\documentclass[aps,twocolumn,floats,prd,nofootinbib,superscriptaddress,10pt]{revtex4-2}

\def\HI{{\rm H\,{\footnotesize I}}}
\def\vcb{v_{cb}}
\def\Ts{T_{\rm S}}
\def\Trad{T_{\rm rad}}
\def\Tb{ T_{{\rm 21}}}
\def\vcb{v_{\rm cb}}
\def\lya{Lyman-$\alpha$~}
\def\mpci{~{\rm Mpc}^{-1}~}

\usepackage{pifont}

\usepackage{amsmath}
\usepackage{amssymb}
\usepackage{graphicx}
\usepackage{multirow}
\usepackage{makecell}
\usepackage{hyperref}
\hypersetup{
	colorlinks=true,
	linkcolor=red,
	filecolor=magenta,      
	urlcolor=blue,
}
\usepackage[normalem]{ulem}
\usepackage[usenames,dvipsnames]{xcolor} 

\begin{document}

	\title{Exploring delaying and heating effects on the 21-cm signature of fuzzy dark matter}

	\author{Debanjan Sarkar}
	\email{debanjan@post.bgu.ac.il}
	\affiliation{Physics Department, Ben-Gurion University of the Negev, Beersheba, Israel}
	\author{Jordan Flitter}%
	\email{jordanf@post.bgu.ac.il }
	\affiliation{Physics Department, Ben-Gurion University of the Negev, Beersheba, Israel}
	\author{Ely D. Kovetz}%
	\email{kovetz@bgu.ac.il }
	\affiliation{Physics Department, Ben-Gurion University of the Negev, Beersheba, Israel}

	\begin{abstract}

	In the fuzzy dark matter (FDM) model, dark matter is composed
	of ultra-light particles with a de Broglie wavelength of $\sim$kpc,
	above which it behaves like cold dark matter (CDM). Due to this, FDM
	suppresses the growth of structure on small scales, which delays the onset of
	the cosmic dawn (CD) and the subsequent epoch of reionization (EoR). 
	This leaves potential signatures in the sky averaged 21-cm signal (global), as well as in the 21-cm fluctuations, which can be sought for with 
	ongoing and future 21-cm global and intensity mapping experiments. 
	To do so reliably, it is crucial to include effects such as the
	dark-matter/baryon relative velocity and Lyman-Werner star-formation feedback, which also act as delaying mechanisms, as well as 
	CMB and \lya heating effects, which can significantly change the amplitude and timing of the signal, depending on the strength of X-ray heating sourced by the remnants of the first stars.
	Here we model the
	21-cm signal in FDM cosmologies across CD and EoR using a modified version of the public code {\tt 21cmvFAST} that accounts for all these additional effects, and is directly interfaced with the Boltzmann code {\tt CLASS} so that degeneracies between cosmological and astrophysical parameters can be fully explored.
	We examine the prospects to distinguish between the CDM and FDM models and forecast 
	joint astrophysical, cosmological and FDM parameter constraints achievable with intensity mapping experiments such as HERA and global signal experiments like EDGES. We find that HERA will be able to detect  FDM particle masses
	up to $m_{\rm FDM}\! \sim \!10^{-19}\,{\rm eV}\!-\!10^{-18}\,{\rm eV}$, depending on foreground assumptions, despite the mitigating effect of the  delaying and heating mechanisms included in the analysis.
		
	\end{abstract}

\maketitle

	\section{Introduction}
	\label{sec:Introduction}
	
	The cold dark matter (CDM) model is a cornerstone of the standard cosmological paradigm.
	It models the dark matter (DM) as a cold, pressureless, non-interacting fluid that dominates the 
	matter budget of the Universe. While CDM has been very successful in explaining the formation 
	and evolution of large scale structure (LSS), the DM particle properties remain elusive. The undetermined small 
	scale behavior of DM, in particular, has been associated with several conflicts between observations and 
	CDM simulations (see Ref.~\cite{DelPopolo:2016emo} for a review).
		
	Fuzzy dark matter (FDM)  is an alternative to CDM~\cite{Hu:2000ke, Hui:2016ltb, Hui:2021tkt}. In this model, DM is composed
	of ultra-light particles with mass as light as $10^{-22}$ eV. FDM thus has a de Broglie wavelength of 
	$\sim\!1\,{\rm kpc}$, 
	below which it features wave-like behavior~\cite{Hui:2016ltb}, which suppresses the growth of structure on small scales. 
	For example, this can be helpful in solving the problem of having too many
	satellite halos in CDM simulations that are not seen in real observations~\cite{Hu:2000ke, Mocz:2019pyf}. 
	FDM also tends to produce cores at the center of halos, rather than infinite cusps like CDM~\cite{Hu:2000ke, Mocz:2019pyf}.
	
	The suppression of structures below a certain scale leads to very interesting astrophysical phenomena
	which can be verified by observations. 
	One such consequence is the delay in the formation of galaxies~\cite{Hu:2000ke}. In hierarchical structure formation, low mass 
	DM halos form early. They subsequently merge as time progresses and form more massive halos.
	These massive halos trap gas which eventually cools and forms luminous structures~\cite{White:1977jf}.
	In the FDM paradigm, halos above the suppression mass cannot collapse before a certain redshift and thus galaxy formation is delayed. 
	Therefore, we can expect this signature of FDM to be found in direct observations of  
	cosmic dawn (CD) and the epoch of reionization (EoR) via the neutral Hydrogen (\HI) 21-cm global signal and its fluctuations~\cite{Lidz:2018fqo}, complementing other FDM probes~\cite{Planck:2019nip,Maleki:2019xya,Sarkar:2021pqh,Unal:2020jiy,DES:2017tss,Bauer:2020zsj,Hlozek:2016lzm,Farren:2021jcd,Porayko:2018sfa,Dentler:2021zij,Church:2018sro,Marsh:2018zyw,Munoz:2019hjh,Hotinli:2021vxg,Blum:2021oxj}.

	A number of experiments have been ongoing or proposed to detect the 21-cm signal. Interferometers like the Hydrogen Epoch of Reionization Array (HERA)~\cite{DeBoer:2016tnn} measure the
	spatial fluctuations in the 21-cm field (see also LOFAR~\cite{Patil:2017zqk}, GMRT~\cite{Paciga:2013fj}, MeerKAT~\cite{Wang:2020lkn} and SKA \cite{Ghara:2016dva}), while the Experiment to Detect the Global EoR Signature (EDGES)~\cite{Bowman:2018yin} and others (e.g.\ SARAS~\cite{Patra:2014qta}, PRIzM~\cite{Chiang:2020pbx}, LEDA~\cite{Bernardi:2017und}) target the sky-averaged signal. They are poised to uniquely probe FDM phenomenology.

	The 21-cm FDM signature has been considered previously, e.g.\ in the context of the claim by the 
	EDGES experiment of a first detection of the global signal from CD~\cite{Lidz:2018fqo,Nebrin:2018vqt}. 
	Recently, Ref.~\cite{Jones:2021mrs} modelled  the FDM impact  on the 21-cm power spectrum and 
	forecasted the expected constraints on the FDM mass from upcoming HERA measurements. They showed that 
	the suppression in the abundance of low-mass halos leads to a delay in the CD and EoR, which 
	strongly impacts the evolution and spatial structure of the 21-cm signal.

		In this work, we revisit the study in Ref.~\cite{Jones:2021mrs} and carry out an analysis of the impact of FDM on the 21-cm signal (including both the global signal and fluctuations), while accounting for several crucial delaying and heating effects which have important implications for this (and virtually any) 21-cm analysis. Moreover, we also study the degeneracies between astrophysical, cosmological and FDM parameters.

Indeed, 21-cm calculations are sensitive to several delaying mechanisms.
First, cosmic structure formation is affected by the relative velocity $\vcb$ between the dark matter and baryons~\cite{Fialkov:2014rba, Barkana:2016nyr, Tseliakhovich:2010bj, Bovy:2012af, Stacy:2010gg, Fialkov:2011iw, Schmidt:2016coo}.  
The baryon acoustic oscillations that occur due to the interaction between the baryon and photon 
fluids before recombination, generate supersonic relative velocities between dark matter and baryons just after
recombination. Ref.~\cite{Tseliakhovich:2010bj} was the first to study the implications of this motion on structure formation 
at high redshifts. It was shown that this supersonic velocity prevents the formation of structures in the 
mini halos ($\sim 10^5-10^6\,{\rm M}_{\odot}$). Subsequent works studied its impact on astrophysics 
and cosmology using various computational tools, including the effects on the evolution of the 21-cm signal \cite{Fialkov:2014rba, Barkana:2016nyr, Bovy:2012af, Stacy:2010gg, Fialkov:2011iw, Schmidt:2016coo}. 
The relative velocity hinders the formation of first stars which then delays the arrival of the CD 21-cm signal. 
Secondly, star formation is also hampered by Lyman-Werner (LW)
radiative feedback~\cite{Fialkov:2012su, Visbal:2014fta, Safranek-Shrader:2012zig, Ricotti:2000at, Haiman:1996rc, Ahn:2008uwe}. The LW photons emitted by each luminous source are absorbed by hydrogen atoms 
as soon as they redshift into one of the Lyman lines of the hydrogen atom. Along the way, 
whenever they hit a LW line they may cause a dissociation of molecular hydrogen.  This, in turn,
applies a negative feedback on star formation, which regulates the process and delays CD.

Our approach is to create a direct interface between the public cosmic microwave background (CMB) Boltzmann code {\tt CLASS}~\cite{Lesgourgues:2011re}
and the public 21-cm code {\tt 21cmFAST}~\cite{Mesinger:2010ne} so that for any model under consideration, 
cosmic evolution is tracked from before recombination and the results are fed as initial conditions to 
generate consistent 21cm realizations.
We use the recent {\tt 21cmvFAST} code~\cite{Munoz:2019rhi} (which accounts for the contribution of molecular cooling halos and both delaying effects above), and recalculate 
for each set of cosmological parameters 
the relative-velocity-dependent quantities that this code uses as input. 
A key advantage of our code, which we plan to make public, is that it enables joint analyses of CMB and 21-cm observations (or mock data) yielding self-consistent and robust cosmological and astrophysical combined parameter constraints.

Furthermore, there are two heating mechanisms 
which may have important impact on raising the intergalactic medium (IGM) temperature
if the poorly-constrained X-ray heating is not extremely efficient,
as we show. These are known as the \lya and CMB heating mechanisms. 
The former mechanism is due to the resonant scattering between \lya photons and the IGM 
atoms~\cite{Chuzhoy:2006au, Chen:2003gc, Oklopcic:2013nda, Ciardi:2009zd}. 
The latter mechanism, recently proposed by Ref.~\cite{Venumadhav:2018uwn}, 
results from the energy transfer from the radio background (which is dominated by the CMB) 
into the IGM, mediated by the \lya photons\footnote{We note that some works debate the significance of this effect~\cite{Meiksin:2021cuh}. Our conclusions are not very sensitive to this, as the \lya heating alone accounts for most of the effect, see below.}.
Following recent literature~\cite{Reis:2021nqf}, we make the necessary modifications to the 21cm code in order to include these effects.
	
While FDM is largely insensitive to the effects of relative velocity and LW feedback, as the suppression scales corresponding to these effects lie well below the suppression scale of FDM with mass of order $10^{-21}\,{\rm eV}$, they strongly affect the baseline CDM signal and hence the ability to distinguish between the two. Meanwhile, we demonstrate that  \lya and CMB heating  can  affect the signal appreciably in both the CDM and FDM scenarios, depending on the dominance of X-ray heating.

Our findings indicate that experiments such as HERA will have the sensitivity to detect FDM with particle mass up to $m_{\rm FDM} \approx 10^{-18}\,{\rm eV}$ in an optimistic foreground scenario and $m_{\rm FDM} \approx 10^{-19}\,{\rm eV}$ in more realistic cases. This bound is roughly an order of magnitude weaker than would be derived without taking into account the delaying and heating mechanisms we focus on.
		Hence this work motivates more careful study of the prospects of the 21-cm signal as a cosmological tool, whether targeting DM or other standard or new physics.	
	
The structure of this paper is as follows. 
In Section~\ref{sec:formalism} we present the formalism used in our calculations, including the \lya and CMB heating which are at the core of our study. In Section~\ref{sec:simulation} we describe the modifications we made to the public code {\tt 21cmvFAST} followed by our prescription for including the new heating effects in the modified code. We present our results in Section~\ref{sec:results} and forecasts with respect to HERA and an EDGES-like experiments in Section~\ref{sec:forecasts}. We conclude in Section~\ref{sec:conclusions}.

	\section{Formalism}
	\label{sec:formalism}
	
	The 21-cm brightness temperature is given by~\cite{Madau:1996cs, Barkana:2000fd, Bharadwaj:2004it}
	\begin{equation}
		\Tb=\frac{\Ts - \Trad}{1+z} \left( 1 - e^{-\tau_{21}}\right) \,,
		\label{eq:T21}
	\end{equation}
	where $\Ts$ is the spin temperature, $\Trad$ is the temperature of the background radiation which is usually
	assumed to be the Cosmic Microwave Background (CMB) with $\Trad = T_{\rm CMB}(z) = 2.726 (1+z) \,{\rm K}$, and $\tau_{21}$
	is the 21-cm optical depth which can be calculated as~~\cite{Pritchard:2008da, Furlanetto:2006jb}
	\begin{equation}
		\tau_{21}=\frac{3hA_{10}c\lambda_{21}^{2}n_{{\HI}}}{32\pi k_{{\rm B}}T_{{\rm S}}(1+z)(dv_{r}/dr)}\,.  
		\label{eq:tau21}
	\end{equation}
	Here, $h$ is the Planck constant, $A_{10}$ is the Einstein $A$-coefficient for the 21-cm emission, $c$ is the speed of light,
	$\lambda_{21}$ is the wavelength of the 21-cm radiation, $n_{\HI}$ is the neutral hydrogen number density, $k_B$ is 
	Boltzmann constant, $dv_{r}/dr$ is the gradient of the comoving velocity along the line of sight. 
	
	The spin temperature can be calculated as~\cite{Pritchard:2008da, Furlanetto:2006jb}
	\begin{equation}
		\Ts=\frac{x_{{\rm rad}} + x_{\alpha} + x_c}{x_{{\rm rad}}T_{{\rm rad}}^{-1} + x_c T_{{\rm K}}^{-1} + 
			x_{\alpha} T_{\rm c,eff}^{-1} }\,,
		\label{eq:Tspin}
	\end{equation}
	where,
	\begin{equation}
		x_{{\rm rad}}=\frac{1-e^{-\tau_{21}}}{\tau_{21}} \,,
		\label{eq:xrad}
	\end{equation}
	$x_{\alpha}$ and $x_c$ are \lya and collisional coupling coefficients respectively, $T_K$ is the IGM kinetic temperature and $T_{\rm c,eff}$
	is the effective color temperature for the \lya radiation.

	The CMB temperature $T_{\rm rad}$ after decoupling simply redshifts with the expansion of the  Universe.
	On the other hand, the evolution of $T_{\rm K}$ depends on several factors and can be described with the following
	equation~\cite{Mesinger:2010ne, Reis:2021nqf}
	\begin{equation}
		\begin{aligned}
			\frac{dT_{{\rm K}}}{dz} & =  2\frac{T_{{\rm K}}}{1+z}+\frac{2T_{{\rm K}}}{3(1+\delta_{b})}\frac{d\delta_{b}}{dz}-\frac{dx_{{\rm e}}}{dz}\frac{T_{{\rm K}}}{1+x_{{\rm e}}}\\
			& -\frac{2}{3k_{{\rm B}}(1+f_{{\rm He}}+x_{{\rm e}})}(\epsilon_{{\rm X}}+\epsilon_{{\rm Compton}}+\epsilon_{{\rm Ly_{\alpha}}}+\epsilon_{{\rm rad}})\,.
		\end{aligned}
		\label{eq:Tk-evol}
	\end{equation}
	The first term in Eq.~\eqref{eq:Tk-evol} corresponds to the Hubble expansion; the second corresponds
	to adiabatic heating and cooling from the structure formation; the third  corresponds to the 
	change in the total number of gas particles due to ionizations; finally the last term corresponds 
	to the heat input from different channels. In our calculations below we consider four major input channels, namely, X-ray heating
	($\epsilon_{{\rm X}}$), Compton scattering ($\epsilon_{{\rm Compton}}$), 
	CMB heating ($\epsilon_{{\rm rad}}$) and \lya heating ($\epsilon_{{\rm Ly_{\alpha}}}$). The different
	input channels are characterized by their respective efficiencies, or rates ($\epsilon_i$).

	The CMB heating rate was calculated in Ref.~\cite{Venumadhav:2018uwn} and is given by
	\begin{equation}
		\epsilon_{{\rm rad}}=\frac{3 x_{{\rm HI}}A_{10}}{4 H(z) (1+z)}x_{{\rm rad}}\left(\frac{T_{{\rm rad}}}{T_{{\rm S}}}-1\right)k_{B}T_{\ast}\,, 
		\label{eq:eps-cmb}
	\end{equation}
	where $T_{\ast}=0.068$ K is the characteristic temperature corresponding to the 21-cm hyperfine transition. Note that $\epsilon_{{\rm rad}}$ is non-zero when $T_{\rm S}$ departs  from $T_{\rm rad}$, that is, when $T_{\rm S}$ 
	has some coupling to $T_{\rm K}$. Ref.~\cite{Venumadhav:2018uwn} showed that this heating has a $\sim\! 10\%$ effect in the absence 
	of X-ray or \lya (or dark matter related) heating, when the background radiation is assumed to be only due to the CMB. In the presence
	of  excess background radiation, this effect can be enhanced, provided other heating mechanisms
	(like X-ray) are not very efficient.

	In order to calculate the \lya heating rate $\epsilon_{{\rm Ly_{\alpha}}}$, one has to solve the steady-state Fokker-Planck equation for obtaining the spectral shapes of the continuum and injected 
	photons~\cite{Chen:2003gc, Chuzhoy:2006au, Oklopcic:2013nda, Ciardi:2009zd}.
	Photons emitted between \lya and Lyman-$\beta$ frequencies (``continuum photons")
	are redshifted to the \lya frequency due to the cosmic expansion and 
	at this point they undergo resonant scattering with \HI, which consequently heats up the IGM. On the other hand,
	photons emitted between the Lyman-$\beta$ and Lyman-limit frequencies
	are absorbed and re-emitted by the higher Lyman-frequencies as they are redshifted. This process creates
	atomic cascades, and eventually the \lya photons 
	produced in these cascades (``injected photons") cool the IGM. 
		
	Estimating the \lya heating is not straightforward. It requires the knowledge of early
	radiative sources, which are largely undetermined mostly due to the lack of observations. 
	The heating rate also depends on the balance between the continuum and injected photons. On average, we expect
	more continuum photons in comparison to  injected photons. The reason is that most of the cascades
	decay to the $2s$ state and produce two photons with frequency smaller than \lya which
	do not contribute to cooling. Only a small fraction of the cascades decay via the $2p$ state and produce
	\lya photons. Ref.~\cite{Chuzhoy:2006au} showed that the gas cannot be heated 
	beyond $\sim\! 100$ K by the \lya photons. Above this temperature, \lya cooling is
	more efficient and it acts to decrease the temperature of IGM.

	\section{Simulation}
	\label{sec:simulation}
	
	We use the \texttt{21cmvFAST}\footnote{\href{https://github.com/JulianBMunoz/21-cmvFAST}{github.com/JulianBMunoz/21-cmvFAST}}
	\cite{Munoz:2019rhi}
	semi-numerical code to generate the observable 21-cm signal. This code is built upon another code, \texttt{21cmFAST}\footnote{\href{https://github.com/andreimesinger/21-cmFAST}{github.com/andreimesinger/21-cmFAST}}
	\cite{Mesinger:2010ne}. 
	\texttt{21cmvFAST} mainly included the effects of DM-baryon relative velocity $\vcb$ and LW radiation 
	feedback into \texttt{21cmFAST}, using pre-calculated input tables of quantities that depend on these effects, given for
	a single set of cosmological parameters (matching Planck cosmology). In order to interface our code with {\tt CLASS} and enable a calculation for any cosmological scenario and any set of input cosmological parameters, we modified the code to calculate all required quantities on the fly. 
		We then added the \lya and CMB heating effects in \texttt{21cmvFAST}, and modified the transfer function from {\tt CLASS} according to the FDM phenomenology, as we describe below.

	\subsection{FDM Transfer function}
	\label{sec:fdm-trans}
	
	The dynamics of structure formation in the FDM model are governed by the non-relativistic 
	Schrödinger–Poisson system of equations. A rigorous solution of this system of equations require
	a lot of computational resource and intricate numerical techniques. However we do not 
	need this rigorous computation for our analysis. We follow Ref.~\cite{Jones:2021mrs} and modify the 
	transfer function as~\cite{Hu:2000ke}
	\begin{equation}
		T^2_{\rm FDM}(k) =  T^2_{\rm CDM}(k) \left[ \frac{\cos(x^3(k))}{1+x^8(k)} \right]^2 \,,
		\label{eq:FDM-transfer}
	\end{equation}
	where $x(k)=1.61 \, [m_{\rm FDM}/e^{-22}\,{\rm eV}]^{1/18} \,\frac{k}{k_{\rm J,eq}}$ and $k_{\rm J,eq}$ is the 
	effective Jeans wavenumber for FDM at  matter-radiation equality, which is given by 
	$k_{\rm J,eq}=9.11\, {\rm Mpc}^{-1} \, [m_{\rm FDM}/e^{-22}\,{\rm eV}]^{1/2}$. Note that this Jeans wavenumber
	depends on the mass $m_{\rm FDM}$ of the FDM particle. As $m_{\rm FDM}$ increases, $k_{\rm J,eq}$ increases and the 
	FDM transfer function approaches the CDM transfer function. We use $m_{\rm FDM}=10^{-21}\,{\rm eV}$ as our fiducial value.
	
	The wavenumber $k_{\rm J,eq}$ defines a characteristic suppression length scale, below which the growth of structures is suppressed. 
	This suppression can also be interpreted as a characteristic mass scale~\cite{Hui:2016ltb}
	that is larger than the minimum halo masses 
	that host the early galaxies during
	CD in the CDM case. As a result, the low-mass halos are suppressed and CD,
	as well as the EoR, is delayed relative to the CDM scenario.

	\subsection{CMB and \lya Heating}
	\label{sec:model-heating}
	
	Accounting for the CMB heating effect requires the knowledge of $x_{{\rm rad}}$. To find it, we solve Eqs.~\eqref{eq:tau21}, \eqref{eq:Tspin} and \eqref{eq:xrad} iteratively, 
	as suggested in Ref.~\cite{Fialkov:2019vnb}, to determine the values of $\Ts$ and $x_{{\rm rad}}$. We start from $x_{{\rm rad}}=1$, and then solve the equations until $\Ts$ and $x_{{\rm rad}}$ converge.
	
	Including the CMB heating in \texttt{21cmvFAST} is straightforward as
	the CMB heating efficiency $\epsilon_{\rm rad}$ depends on the local values of $T_{\rm S}$, $T_{\rm rad}$, $x_{\HI}$
	and $x_{\rm rad}$. In \texttt{21cmvFAST}, the whole simulation box is divided into a number of finite grids and
	the calculation of the different fields (like $T_{\rm S}$, $T_{\rm rad}$, $x_{\HI}$, {\it etc.}) is done on these grids. Hence, the calculation of $\epsilon_{{\rm rad}}$ only required us to implement Eq.~\eqref{eq:eps-cmb}.
		
	For incorporating the \lya heating mechanism in \texttt{21cmvFAST}, we follow closely the prescription given in Ref.~\cite{Reis:2021nqf}, but without their multiple scattering scheme. According to that prescription, the \lya heating rate is proportional to the \lya flux. Besides the usual stellar contribution to the \lya flux (which we assume it entirely comes from the Population-II stars),
	\texttt{21cmvFAST} also takes into
	account the production of  \lya photons by the X-ray excitation of hydrogen atoms. This
	contribution is actually added to the \lya photon intensity to calculate the \lya 
	coupling $x_{\alpha}$. However, we do not incorporate this contribution while calculating the \lya 
	heating, for two reasons: $(i)$ for low X-ray efficiency, this contribution is negligible
	($\leqslant 1-2\%$), $(ii)$ for high X-ray efficiency, although this contribution can be 
	$\approx10-20\%$ or higher, the overall \lya heating effect is not very significant~\cite{Reis:2021nqf}.
	Therefore, the contribution of \lya photons from X-ray excitation can be safely ignored for
	the calculation of the \lya heating.
	
	We mentioned the difference between the continuum and injected photons in Section~\ref{sec:formalism}.
	In the simulation, we separately calculate the continuum and injected \lya photon intensities
	that are used to calculate the corresponding heating efficiencies. 
	These efficiencies depend on the local values of $T_{\rm K}$, $T_{\rm S}$ and $\tau_{\rm GP}$ (the Gunn-Peterson optical depth). As for
	 CMB heating, we also calculate the efficiencies on each simulation grid and then add the contributions
	to the evolution equation of $T_{\rm K}$ (Eq.~\eqref{eq:Tk-evol}).
	
	Note that
	the calculation of the \lya heating efficiencies increases the overall run-time of the code as it requires performing double integration at each voxel of the simulation. To minimize
	the run-time, we calculate the efficiencies as functions of $T_{\rm K}$, $T_{\rm S}$ and $\tau_{\rm GP}$ 
	separately and save those as tables. We then interpolate the efficiencies on the grids
	while the full simulation is running. We have checked that our interpolation scheme yields the (almost) same heating efficiencies when they are calculated locally at each grid point.

	\subsection{Model \& simulation parameters}
	
	The \texttt{21cmvFAST}~\cite{Munoz:2019rhi} code uses a number of astrophysical and cosmological parameters. 
	The astrophysical
	parameters are: $\zeta$ (describes the efficiency of ionizing photon production),
	$\lambda_{\rm MFP}$ (mean free path of the ionizing photon),
	$V^{(0)}_{\rm cool}$ (minimum halo mass for  molecular cooling in the absence of relative velocity),
	$V^{\HI}_{\rm cool}$ (minimum halo mass for atomic cooling),
	$\log_{10} \left( L_{\rm X}/{\rm SFR} \right)$ (log of X-ray luminosity, normalized by the star formation rate SFR, in units of $\mathrm{erg}\,\,\mathrm{s}^{-1}\,\,M_\odot^{-1}\,\,\mathrm{yr}$),
	$\alpha_X$ (X-ray spectral index),
	$f^0_{\ast}$ (fraction of baryons in stars),
	$E_{\rm min}$ (threshold energy, below which we assume all X-rays are self-absorbed near the sources).
	
	We assume a flat Universe with the following cosmological parameters: 
	$h$ (Hubble parameter),
	$\sigma_{8,0}$ (standard deviation of the current matter fluctuation smoothed at scale $8\,h^{-1}$Mpc)
	$\Omega_{m0}$ (total matter density at present),
	$\Omega_{b0}$ (total baryon density at present),
	$n_s$ (spectral index of the primordial power spectrum),
	$T_\mathrm{CMB}$ (current CMB temperature). The fiducial values of these parameters are given in 
	Table~\ref{tab:parameters-fid}.

	We run our modified version of \texttt{21cmvFAST} with box sizes $600$ Mpc and $1$ Mpc resolution to compute the 21-cm global
	signal and fluctuations. 
		We checked that the choice of a $600$ Mpc box retains
	sufficient $\vcb$ power at large scales and the power spectra show good convergence with a $900$ Mpc box
	results.

	\begin{table}[]
		\begin{tabular}{|c|c|}
			\hline
			Parameters & Fiducial Values \\  \Xhline{4\arrayrulewidth}
			$\zeta$ & $20$    \\ \hline
			$\lambda_{\rm MFP}$ & $15$ Mpc  \\ \hline
			$V^{(0)}_{\rm cool}\, [{\rm km/s}]$ & 4  \\ \hline
			$V^{\HI}_{\rm cool}\, [{\rm km/s}]$ & 17  \\ \hline
			$\log_{10} \left( L_{\rm X}/{\rm SFR} \right)$ & $38,39,40$  \\ \hline
			$\alpha_X$ & $1.2$  \\ \hline
			$f^0_{\ast}$ & $0.05$  \\ \hline
			$E_{\rm min}$ & $0.2$ keV \\  \Xhline{4\arrayrulewidth}
			$\sigma_{8,0}$ & 0.8102  \\ \hline
			$h$ & 0.6766   \\ \hline
			$\Omega_{m0}$ & 0.3111   \\ \hline
			$\Omega_{b0}$ & 0.0489  \\ \hline
			$n_s$ & 0.9665  \\ \hline
			$T_{\rm CMB}$ & 2.7255  \\  \Xhline{4\arrayrulewidth}
			$m_{\rm FDM}$ & $10^{-21}\,{\rm eV}$ \\ \hline
					
		\end{tabular}
		\caption{Our main simulation parameters and their fiducial values.}
		\label{tab:parameters-fid}
	\end{table}

	In our simulations, we  consider three different X-ray heating efficiencies: 
	$\log_{10}(L_{\rm X}/{\rm SFR})=38$ (low X-ray efficiency), $39$ (moderate efficiency),
	and $40$ (high efficiency).
	We  include the effects of $\vcb$ and LW radiation feedback in our simulations, exploring  the latter for three cases,
	(i) no feedback, (ii) low feedback and (iii) regular feedback, as defined in Ref.~\cite{Munoz:2019rhi}.
	Meanwhile, we also consider both CMB and \lya heating.  
	Combining all the different parameters and effects, for both 
	CDM and FDM scenarios, results in a large number of simulations. 
	To mitigate this, we do not discuss the effects of CMB and \lya heating separately, but 
	rather combine them, referring to the sum as ``additional heating". 
	
	\section{Results}
	\label{sec:results}
	
	\subsection{Delaying and Heating Effects}
	
	\begin{figure}
		\includegraphics[width=\columnwidth]{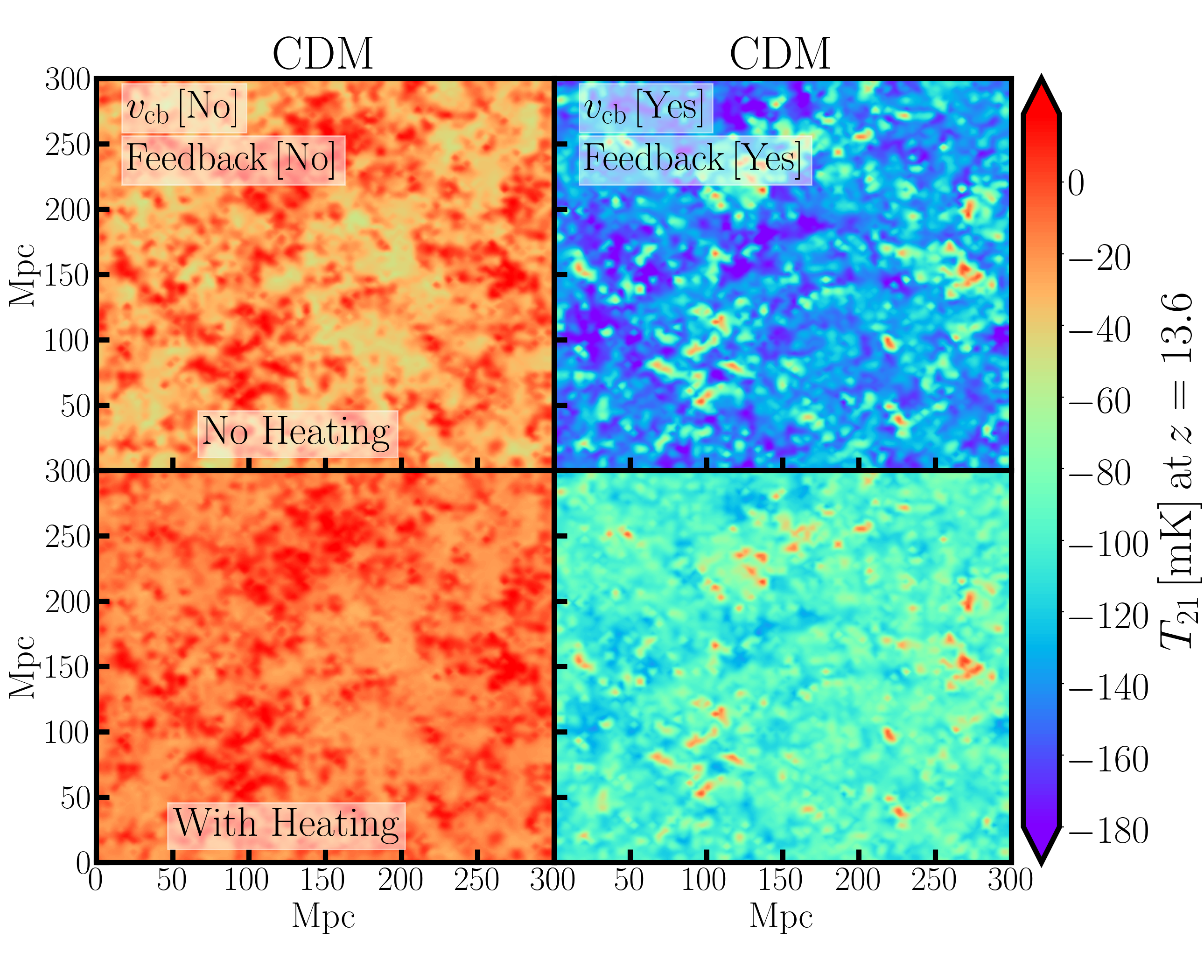}
		\caption{The spatial fluctuations of 21-cm brightness temperature  $T_{21}$ 
			(Eq.~\ref{eq:T21}) at $z\!=\!13.6$.  Top and bottom panels show results without and
			with additional heating effects, respectively. The left column corresponds to CDM model without
			the relative velocity $\vcb$ and LW feedback, whereas the right column 
			corresponds to CDM model with both  the relative velocity $\vcb$ and LW feedback effects.
			Each result here is obtained from a simulation 
			slice $300$ Mpc in length and $3$ Mpc in thickness. We use $\log_{10}(L_{\rm X}/{\rm SFR})\!=\!39$
			for all the simulations.}
			
		\label{fig:21cm_fluctuations}
	\end{figure}
	
	In Fig.~\ref{fig:21cm_fluctuations}, we show the spatial fluctuations of
	the 21-cm brightness temperature $T_{21}$ (Eq.~\ref{eq:T21}) for CDM 
	at $z=13.6$. The dip of the global signal $\langle T_{21} \rangle$
	for the FDM models is very close to this redshift. The top panels show results without 
	the additional heating and the bottom panels show results with additional heating. Considering
	the top panels, we see that for CDM without $\vcb$ and LW feedback, the $T_{21}$ values 
	lie in the range $-50 \, {\rm mK} \, \lesssim T_{21} \lesssim 20 \, {\rm mK} \,$
	and the average temperature $\langle T_{21} \rangle$ $\approx -24$ mK. 
	When we include additional heating, the average temperature of the box rises to
	$\langle T_{21} \rangle$ $\approx -13$ mK and now
	the $T_{21}$ values lie in the range $-20 \, {\rm mK} \, \lesssim T_{21} \lesssim 20 \, {\rm mK} \,$. 
	With $\vcb$ and LW feedback, the $T_{21}$ values without (with) additional heating
	lie in range $-160 \, {\rm mK} \, \lesssim T_{21} \lesssim -10 \, {\rm mK} \,$ 
	($-120 \, {\rm mK} \, \lesssim T_{21} \lesssim -10 \, {\rm mK} \,$) with average value 
	$\langle T_{21} \rangle$ $\approx -135$ mK ($\approx -100$ mK). 
	
\begin{figure*}
		\centering
		\includegraphics[width=\textwidth]{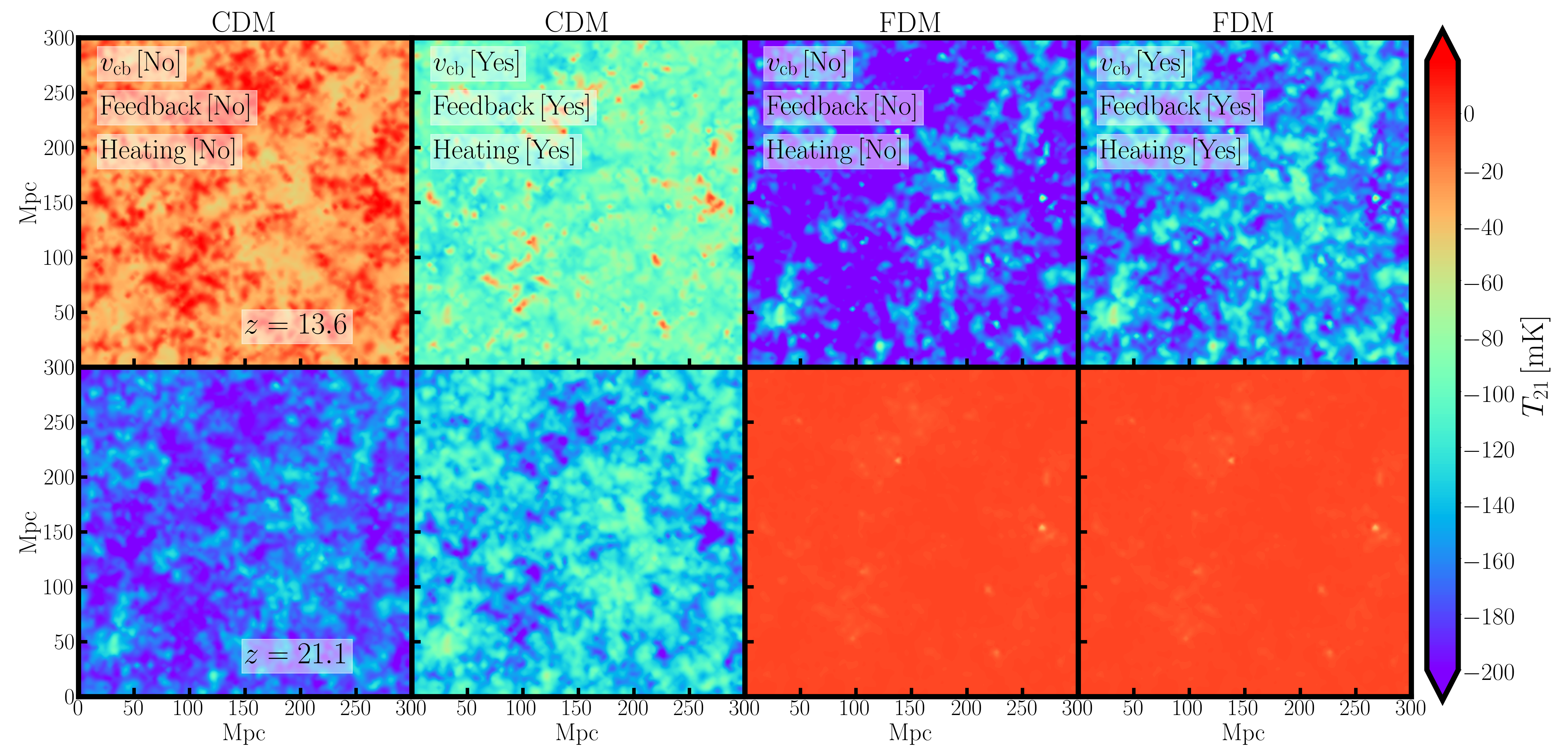}
		\caption{The spatial fluctuations of the 21-cm brightness temperature  $T_{21}$ 
			(Eq.~\eqref{eq:T21}) at $z\!=\!13.6$ and $z\!=\!21.1$. The odd and even columns show results without and
			with additional (\lya and CMB) heating and delaying (relative velocity $\vcb$ and LW feedback) effects, respectively. 
			The first two columns show results for CDM and the last two for FDM. 
			Each result here is obtained from a simulation 
			slice $300\,{\rm Mpc}$ in length and $3\,{\rm Mpc}$ in thickness, and setting $\log_{10}(L_{\rm X}/{\rm SFR})\!=\!39$.}
		\label{fig:21cm_fluctuations_FDM}
	\end{figure*}

	In Fig.~\ref{fig:21cm_fluctuations_FDM}, we compare between the CDM and FDM models at two redshifts, 
	$z=13.6$ and $z=21.1$. At $z=21.1$, both \lya coupling and additional heating are present for CDM,
	whereas both of these are yet to start for FDM. We find that almost all the pixels of both the FDM boxes 
	show $T_{21}\approx 0$ mK, and there is no visible difference when we either consider or drop the effects
	of $\vcb$ and feedback. On the other hand, the effects of $\vcb$, feedback and additional heating are
	very apparent for the CDM boxes. Without (with) all these effects, the $T_{21}$ values lie in range
	$-200 \, {\rm mK} \, \lesssim T_{21} \lesssim -110 \, {\rm mK} \,$
	($-200 \, {\rm mK} \, \lesssim T_{21} \lesssim -80 \, {\rm mK} \,$) with 
	$\langle T_{21} \rangle$ $\approx -170$ mK ($\langle T_{21} \rangle$ $\approx -100$ mK).
	Considering FDM results at $z=13.6$, we find that 
	without additional heating, most of the $T_{21}$ value are below $-200$ mK and a few pixels 
	show values around $-100$ mK, with an average $\langle T_{21} \rangle$ $\approx -180$ mK. When
	the additional heating is included, the average temperature rises to
	$\langle T_{21} \rangle$ $\approx -150$ mK. Although, the $T_{21}$ values still lie in the range
	$-200 \, {\rm mK} \, \lesssim T_{21} \lesssim -100 \, {\rm mK} \,$, more $T_{21}$ values are
	now close to $-100$ mK which increases the overall average. Note that the highest 
	peak values of $T_{21}$, where we expect the sources to lie, do not change by much when we 
	include additional heating. Only the low $T_{21}$ regions around the highest peaks
	show increased temperature with additional heating.
	
	From the discussion above, we conclude that
	both the CDM and FDM models are affected by the additional heating.
	Overall, additional heating alters the spatial structure of the 21-cm fluctuations. 
	Also, $\vcb$ and LW feedback have no visible effects on $T_{21}$ fluctuations for FDM. 
	We will quantify both delaying and heating effects through the 21-cm global signal and power spectrum.

	\subsection{Global Signal}
	
	\begin{figure}
		\centering
		\includegraphics[width=\columnwidth]{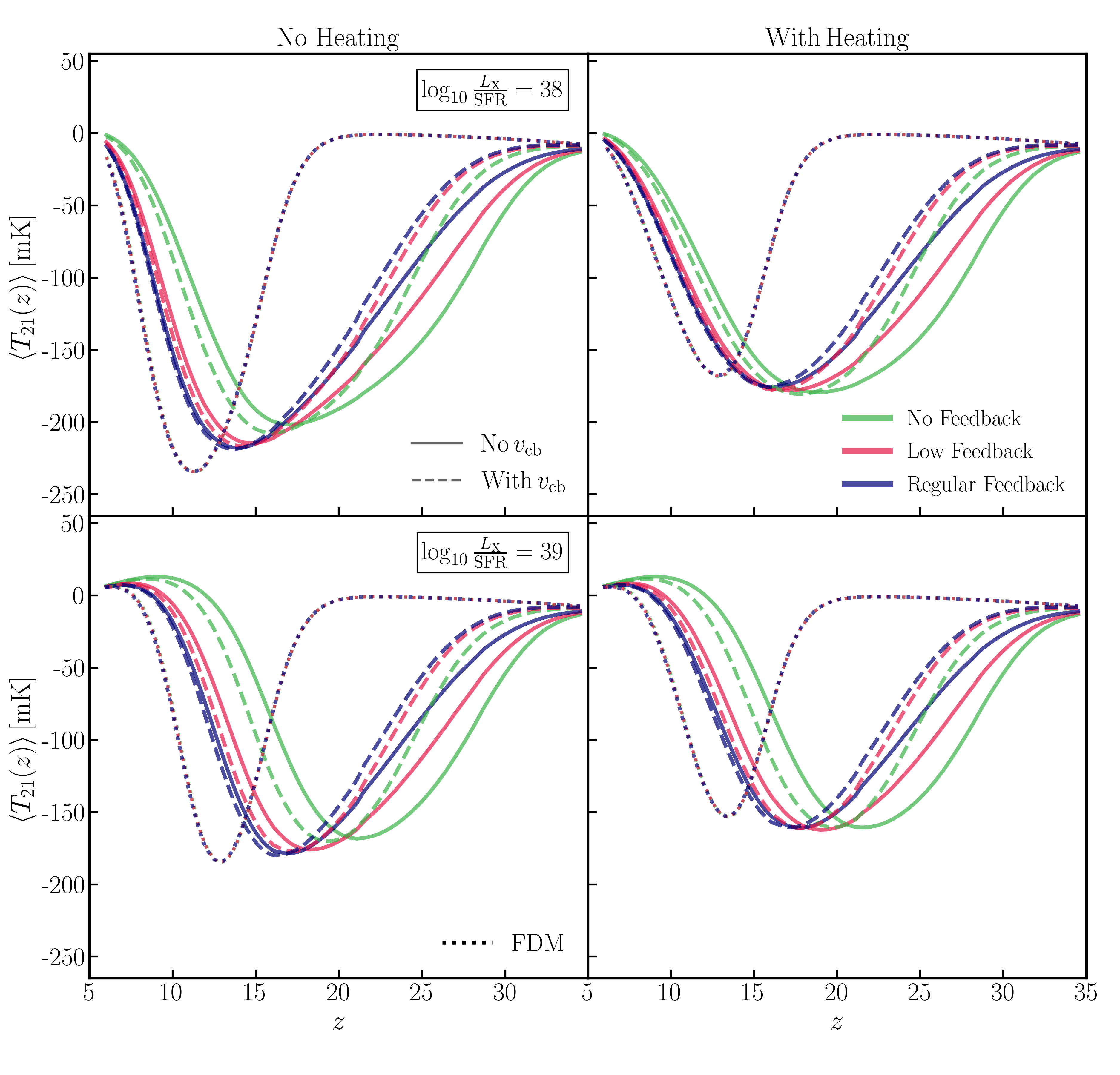}
		\caption{The 21-cm global signals $\langle T_{21}(z) \rangle$ as a function of 
			redshift $z$. The top panels correspond to low X-ray efficiency ($\log_{10}(L_{\rm X}/{\rm SFR})=38$),
			and bottom panels correspond to moderate X-ray efficiency ($\log_{10}(L_{\rm X}/{\rm SFR})=39$).
			Solid and dashed curves refer to the CDM models without and with the relative velocity $\vcb$ effects, whereas the
			different colors indicate  different LW feedback strengths.  As relative velocity $\vcb$ does not affect the FDM results, 
			we show those with dotted curves where the different colors indicate different LW feedback strengths. Note that 
			for FDM, the curves with different LW feedback strengths overlap. This again indicates that LW feedback has no effect
			on FDM. 
		} 
		
		\label{fig:global-heating}
	\end{figure}
	
	\begin{figure}
		\centering
		\includegraphics[width=\columnwidth]{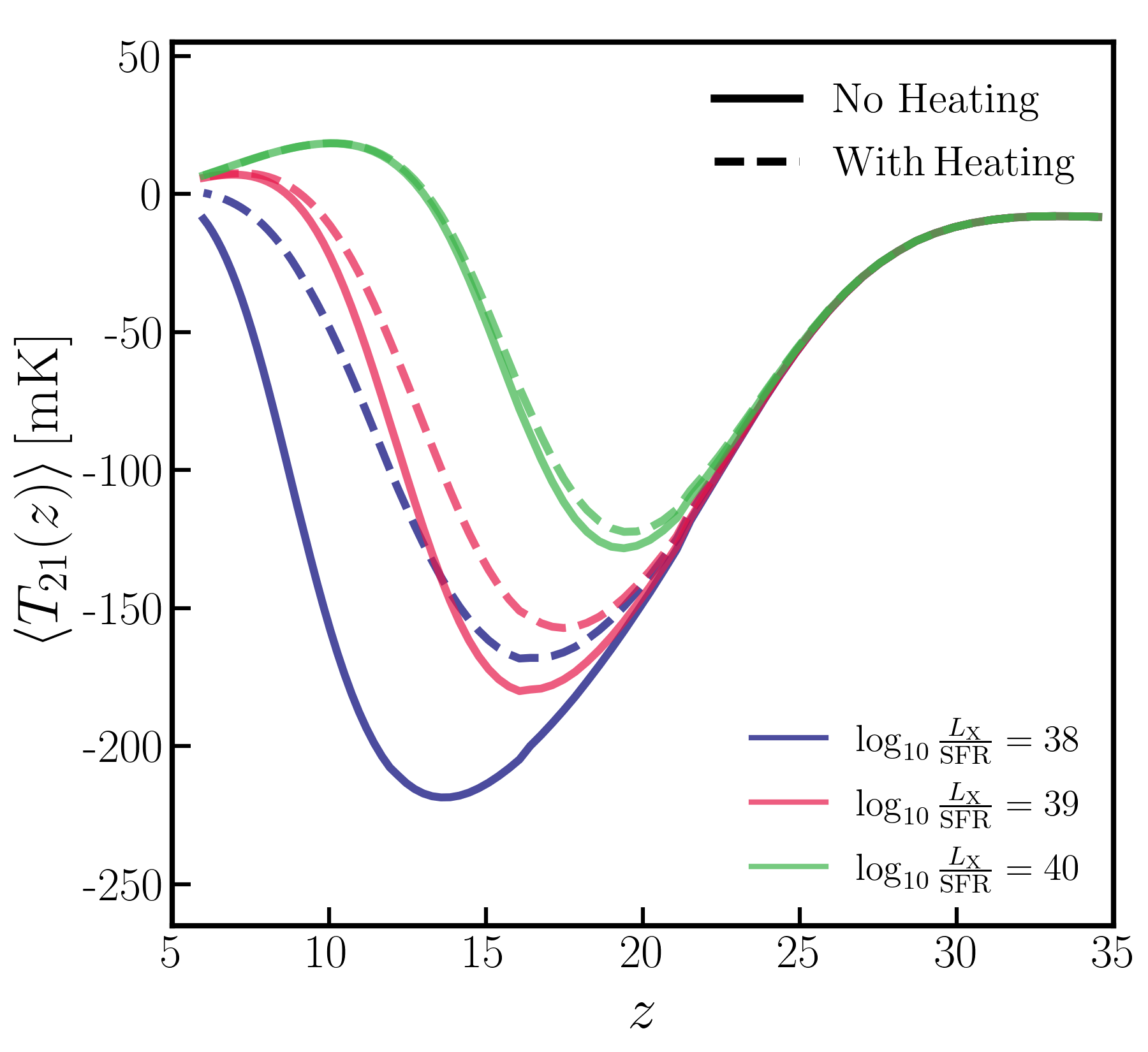}
		\caption{The 21-cm global signals $\langle T_{21}(z) \rangle$ as a function of 
			redshift $z$ for three different X-ray heating efficiencies: low  ($\log_{10}(L_{\rm X}/{\rm SFR})=38$),
			moderate  ($\log_{10}(L_{\rm X}/{\rm SFR})=39$) and high  ($\log_{10}(L_{\rm X}/{\rm SFR})=40$).
			All  curves are for CDM models without $\vcb$ and feedback.
			The solid and dashed curves correspond to models without and with additional heating effect.
		}
		\label{fig:global-xray}
	\end{figure}
	
	Fig.~\ref{fig:global-heating}
	shows the global signal for both CDM and FDM models with two different X-ray efficiencies,
	$\log_{10}(L_{\rm X}/{\rm SFR})=38$ and $39$.
	The overall shape and the minimum value of the signal depend on the various effects
	like $\vcb$, LW feedback and additional heating.
	
	We discuss the different cases one by one. We first consider the top-left panel which shows the global
	signals for $\log_{10}(L_{\rm X}/{\rm SFR})=38$ (low X-ray efficiency) and without additional heating. 
	For CDM, without LW feedback and $\vcb$, we see that the minimum occurs at $z\sim17$. 
	However, when the LW feedback or $\vcb$ is considered, the minimum is shifted towards smaller 
	redshift. Both $\vcb$ and LW feedback prevent the formation of luminous structures inside small halos and
	 increase the mass of the smallest halos that can form stars. This process delays the beginning
	of the \lya coupling era and the minimum of the global signal
	is shifted to smaller redshift. In presence of both LW feedback and $\vcb$, 
	this effect is strongest and the minima for the CDM model occur at the smallest redshift ($z\sim13$).
	Moreover, the shape of the signal depends
	very much on the presence of LW feedback and $\vcb$.
	
	In contrast, for the FDM model we see that the shape and the minimum of the
	global signal do not depend on neither LW feedback nor $\vcb$. This is expected, as the length scale 
	below which the halos are suppressed in FDM model is well above the effective Jeans scales of both $\vcb$ and 
	LW feedback, i.e.\ the minimum halo mass that can contain first galaxies in FDM paradigm
	is well above the minimum halo masses that are affected by the $\vcb$ and LW feedback. 
	In the $z$ range of Fig.~\ref{fig:global-heating}, 
	the kinetic temperature drops off as $(1+z)^2$. Therefore, the amplitude of the minimum of the global
	signal depends on when the  \lya coupling saturates and $T_{\rm S}$ couples with $T_{\rm K}$. In the CDM
	model, $\vcb$ and LW feedback delays the star formation, so we get a minimum signal when both 
	$\vcb$ and LW feedback are present. Similarly, FDM shows the minimum temperature in comparison to CDM. 
	
	However, the above is true when X-ray heating is not very efficient and additional heating sources
	are not present. In the presence of additional heating (top-right panel), 
	or efficient X-ray heating (bottom-left panel), we see that the minimum of the signal 
	(in comparison to the top-left panel) occurs at slightly higher redshift. This happens because the 
	external heating sources tend to increase $T_{\rm K}$ before the signal reaches the minimum value where
	the \lya coupling saturates. This effect is maximal when all the heating sources come into play. However,
	the additional heating is no longer important when we have very efficient X-ray heating. 
	This we can see in the bottom panels, {\it i.e.} for $\log_{10}(L_{\rm X}/{\rm SFR})=39$. Here, 
	the curves with (right panel) and without (left panel) additional heating are not very different. 
	
	In order to demonstrate this
	more clearly, we have plotted the CDM results for three different X-ray
	heating efficiencies in Fig.~\ref{fig:global-xray}. We see that the solid (without additional heating)
	and dashed (with additional heating) curves almost overlap for $\log_{10}(L_{\rm X}/{\rm SFR})=40$. This happens 
	when X-ray heating becomes so efficient that it comes into play even before the CMB or \lya heating
	start to act. Another important difference between the CDM and FDM models is that the 
	additional heating effect is maximal for the FDM models. 
	Note that, between CMB and \lya heating, the former is generally more efficient above a certain redshift and it is sufficient to explain this 
	difference in terms of CMB heating. This is not necessarily correct when multiple scatterings are considered, which we leave to future work.
	The CMB heating is more efficient if $(T_{\rm rad}/T_{\rm S} -1)$ is larger and $x_{\HI}$ is higher. 
	Both of these are true for FDM models in comparison to CDM, 
	which is why the additional heating is more prominent for the FDM models.

	\begin{figure*}
		\centering
		\includegraphics[width=1\textwidth]{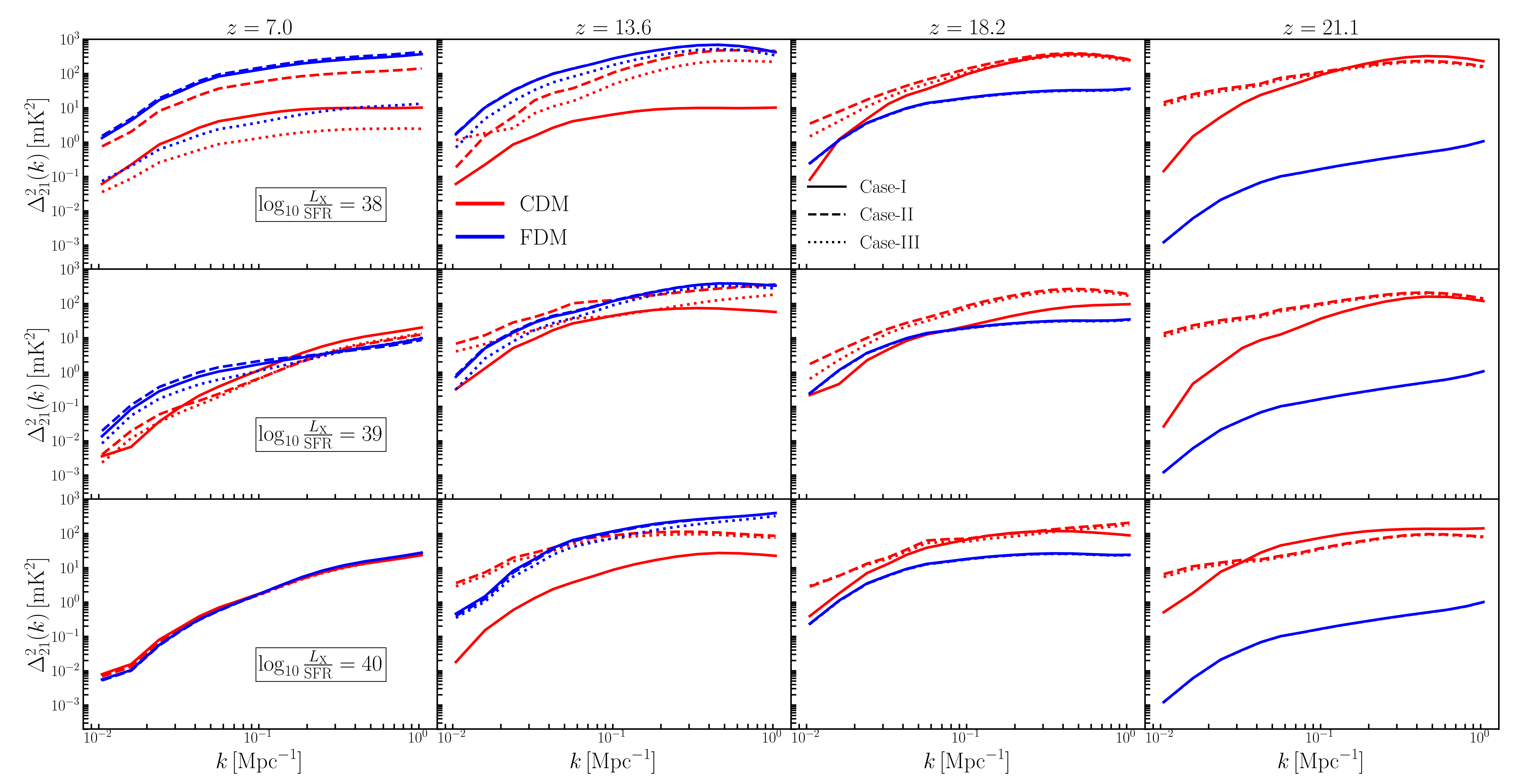}
		\caption{The 21-cm power spectrum $\Delta^2_{21}(k)$ (Eq.~\ref{eq:power-spec}) as a function 
			of the wave vector $k$. The top, middle and bottom panels show results for 
			three different X-ray heating efficiencies: low  ($\log_{10}(L_{\rm X}/{\rm SFR})=38$),
			moderate  ($\log_{10}(L_{\rm X}/{\rm SFR})=39$) and high  ($\log_{10}(L_{\rm X}/{\rm SFR})=40$, respectively.
			Red and blue colors indicate CDM and FDM models, respectively.  Solid, dashed and dotted curves 
			refer to 	\texttt{Case-I} (no $\vcb$, no LW feedback, no additional heating),
			\texttt{Case-II} (with $\vcb$, with regular LW feedback, no additional heating) and 
			\texttt{Case-III} (with $\vcb$, with regular LW feedback, with additional heating), respectively.
			Each column corresponds to a particular redshift which is indicated at the top.
		}
		\label{fig:power-spectrum-k}
	\end{figure*}
	
	\subsection{Fluctuation Power Spectrum}
	In this section, we discuss the 21-cm power-spectrum
	which is defined as
	\begin{equation}
		\Delta^2_{21}(k)=\frac{k^3P_{21}(k)}{2 \pi^2} \,[{\rm mK}^2],
		\label{eq:power-spec}
	\end{equation}
	where $P_{21}(k)=\langle \Tilde{T}_{21}(k) \Tilde{T}_{21}^{\ast}(k) \rangle$ and $\Tilde{T}_{21}(k)$ is the 
	Fourier transform of $T_{21}-\langle T_{21}\rangle$. Fig.~\ref{fig:power-spectrum-k} shows the 21-cm power spectra
	as functions of $k$
	for three different X-ray efficiencies, $\log_{10}(L_{\rm X}/{\rm SFR})=38,39,40$. 
	We show our results at four different redshifts: 
	$(i)$ $z=21.1$: which is within CD for the CDM models,
	$(ii)$ $z=18.2$: where we expect the additional heating effects to start for the CDM models,
	$(iii)$ $z=13.6$: very close to the dip of the global signal for the FDM models,
	$(iv)$ $z=7$: towards the end of reionization where $x_{\HI}$ drops and we expect  X-rays to dominate for
	very efficient X-ray heating. We see that the power spectra show very different scale dependence at these redshifts
	for the different models. We have chosen three cases to demonstrate different effects, 
	\texttt{Case-I}: no $\vcb$, no LW feedback, no additional heating,
	\texttt{Case-II}: with $\vcb$, with Regular LW feedback but with no additional heating and 
	\texttt{Case-III}: with $\vcb$, with Regular LW feedback, and with additional heating.

	We first consider the top panels, i.e.\
	power spectra for low X-ray efficiency $\log_{10}(L_{\rm X}/{\rm SFR})=38$. At $z=21.1$, we do not expect the X-ray
	or even additional heating effects to come into play. Therefore, the shape and amplitude of the 
	power spectra here depend on the onset of the \lya coupling in the different models. The \lya coupling starts
	early for the CDM models in comparison to FDM models.  In absence of \lya coupling,
	the $T_{\rm S}$ in FDM models is still close to $T_{\rm rad}$. Due to this, 
	we see that the amplitude of the 21-cm power spectra is very low (factor of $>100$ smaller) 
	for the FDM models in comparison to the CDM models. Among the CDM models, relative velocity and LW feedback 
	cut out smaller mass halos and the higher mass halos that remain are highly biased. This enhances the 
	large scale (small $k$) amplitude of the 21-cm power spectrum for \texttt{Case-II} and \texttt{III} in
	comparison to \texttt{Case-I}. 
	At $z=18.2$, the \lya coupling has already started for the FDM model, $T_{\rm S}$ gets closer to $T_{\rm K}$
	and the amplitude of the power spectrum increases for FDM. However, heating is still not started for the
	FDM models at this redshift. For the CDM models, $\vcb$ and LW feedback actually delay the heating, and heating
	reduces the amplitude of the 21-cm power spectrum as it takes the $T_{\rm K}$ towards $T_{\rm rad}$. This can be seen
	in this panel as the curve with heating (dotted red) lies below the curve with no heating. 
	
	At $13.6$, the FDM models are very close to the dip of the global signal. This implies that the fluctuations 
	are maximally away from $T_{\rm rad}$ and 21-cm fluctuations show maximum power. However, FDM models with
	heating show a smaller amplitude for the power spectrum. For the CDM models, we see a minimum amplitude for 
	\texttt{Case-I}
	power spectrum and a maximum amplitude for \texttt{Case-II} power spectrum. Here also, heating reduces the amplitude 
	and \texttt{Case-III}
	remains below \texttt{Case-II}. At $z=7$, we see similar features as we see in $z=13.6$, 
	only the power in \texttt{Case-III} is smallest 
	here. If we consider the middle and bottom panels, which means if we increase the X-ray heating efficiency, we see
	that the difference between \texttt{Case-II} and \texttt{Case-III} becomes small for both CDM and FDM. 
	This again shows that
	the additional heating is not important if the X-ray heating is highly efficient. For $\log_{10}(L_{\rm X}/{\rm SFR})=40$,
	we see that all the curves almost overlap at $z=7$ which marks the end of the reionization stage. By this time, X-ray
	heating dominates over the other physical effects and we see almost no difference between
	CDM and FDM models.

	\begin{figure}
		\centering
		\includegraphics[width=\columnwidth]{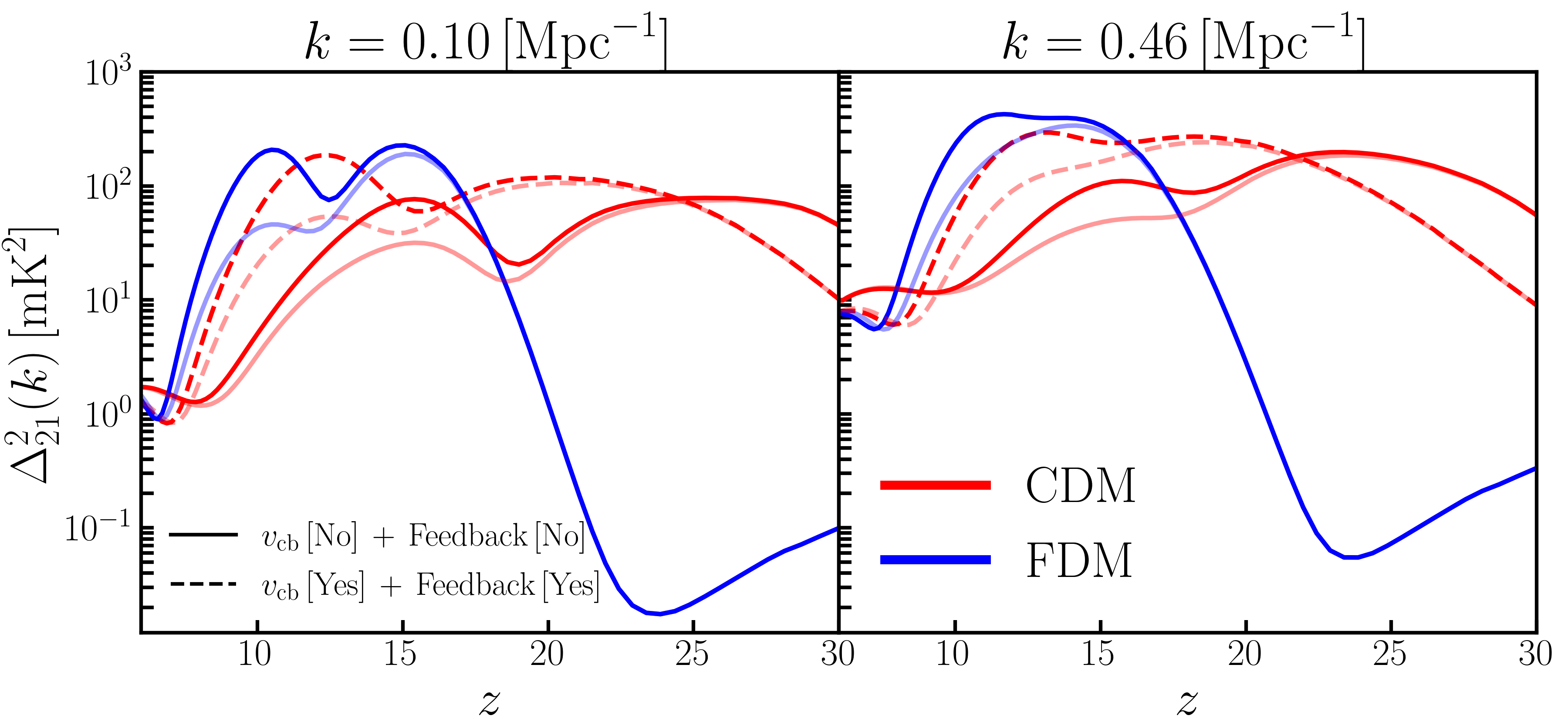}
		\caption{The redshift $z$ evolution of the 21-cm power spectrum $\Delta^2_{21}$ at 
			$k=0.1\mpci$ and $k=0.46\mpci$. The red and blue colors, respectively, indicate CDM and FDM models,
			whereas the deep and light shades  of those indicate models without and with additional heating.
			Solid curves show results without the relative velocity $\vcb$ and LW feedback, and dashed curves show
			results with both relative velocity $\vcb$ and LW feedback. Here the X-ray heating efficiency is 
			kept fixed  at $\log_{10}(L_{\rm X}/{\rm SFR})=39$ for all the curves. 
		}
		\label{fig:power-spectrum-z}
	\end{figure}

	In Fig.~\ref{fig:power-spectrum-z}, we show the redshift dependence of the power spectrum amplitude 
	at two specific $k$ values, for completeness. We first consider the \texttt{Case-I} for CDM at $k=0.1\mpci$. 
	We can see three distinct epochs where three different effects dominate the fluctuation fields.
	Fluctuations in \lya coupling dominate at $z>20$ and this contribution peaks around $z\sim24$ when $x_{\alpha}$
	becomes $1$. The dip around $z\sim20$ marks the transition from the \lya to X-ray heating domination. 
	X-ray heating fluctuations dominate inbetween $10<z<20$ and this peaks around $z\sim 16$. Fluctuations due
	to X-ray heating reduce and show a dip at $z\sim10$. Below this redshift, the ionization fluctuations during the
	reionization epoch dominate and the power spectrum again rises. 
	
	At even smaller redshifts ($z<6$, which we do not show here), the power spectrum eventually 
	goes down due to a rapid decline in the neutral fraction $x_{\HI}$. These three distinct epochs are 
	clearly visible in all other models, as well as at $k=0.46\mpci$. Now, when we consider \texttt{Case-II} for CDM, 
	i.e.\ we include the effects of $\vcb$ and LW feedback, we see that all the features (peaks and dips) are 
	shifted toward smaller redshift by roughly $\Delta z \sim 4$. In the range $8<z<25$, 
	the amplitude of the power spectrum is also
	higher in this case. This again verifies the fact that 
	$\vcb$ and LW feedback delay the CD and therefore all the successive epochs. 
	
	Comparing the FDM models with CDM, we see a similar effect.
	The absence of smaller mass halos in FDM delays the onset of the domination of \lya, X-ray and ionization
	fluctuations, and the corresponding epochs are shifted by $\Delta z \sim 8, 5$ and $2$ respectively. 
	Remember that $\vcb$ and LW feedback have no visible effects for FDM models. In the range $7<z<19$, the
	amplitude of the FDM power spectrum is higher than almost all of the CDM models. Now, if we 
	include additional heating, we see that the amplitude of the power spectrum, compared to the 
	no heating cases, decreases below the peak
	redshift of the \lya epoch for all the models and this remains true until the end of the X-ray heating 
	epoch. This suppression in power is maximal near the peak of the X-ray heating era, and this effect is 
	more prominent for the FDM models. The additional heating decreases the amplitude of the power spectrum 
	as it increases  $T_{\rm K} \approx T_{\rm S}$ and overall decreases the contrast $(T_{\rm rad}-T_{\rm S})$. 
	The additional heating also shifts the peaks and the dips, but only very slightly. 
	Considering the right panel, we observe that the above
	discussion is qualitatively true for $k=0.46\mpci$ as well.

	\section{Forecast with HERA}\label{sec:forecasts}
	
	\subsection{Sensitivity Calculation}
	\label{sec:sensitivity}
	
	In this section, we discuss the possibility of measuring the 21-cm power spectrum using the upcoming
	HERA 21-cm intensity mapping experiment~\cite{DeBoer:2016tnn}. HERA is located in the Karoo Desert of South Africa and is designed to measure the 21-cm fluctuations 
	from CD ($50$ MHz or $z\sim27$) to the reionization era ($225$ MHz or $z\sim5$). 
	The final stage of HERA is expected to have $350$
	antenna dishes, each with a diameter of $14$ m. Out of the $350$ dishes, $320$ will be 
	placed in a close-packed hexagonal configuration and the remaining $30$ will be placed at longer 
	baselines. HERA will mainly operate as a drift scan telescope where the telescope will 
	point toward the zenith and the scanning will be done as the Earth rotates. 
	
	We calculate the sensitivity of HERA using the publicly available
	package \texttt{21cmSense}\footnote{\href{https://github.com/jpober/21cmSense}{github.com/jpober/21cmSense}}
	~\cite{pober12, Pober:2013jna}. 
	This code accounts for the $u-v$ sensitivities of each antenna
	in the array, and calculates the possible errors in the 21-cm power spectrum measurement,
	including cosmic variance. The \texttt{21cmSense} package assumes a receiver 
	temperature of $100$ K and a sky temperature of $T = 60\, {\rm K}\, (\nu / 300 \,{\rm MHz})^{-2.55}$,
	and the combination of both is called the system temperature $T_{\rm sys}$. We assume a total 
	observing time of $1080$ hours to calculate the sensitivity of HERA. The measurements of the
	power spectrum are assumed to be done in bandwidths of $8\,{\rm MHz}$ and simultaneously across 
	the redshift range $z=6$ to $27$. Although these simultaneous measurements across a large $z$
	range are practically impossible~\cite{Ewall-Wice:2015uul}, this assumption will provide us some useful insight
	which is sufficient for this analysis. 
	
	The real challenge in 21-cm observations are the Galactic and extra-galactic foregrounds
	which plague the tiny 21-cm signal~\cite{Bowman:2008mk, Dillon:2012wx, Hazelton:2013xu, Liu:2011hh}. 
	However, the foregrounds are expected to be spectrally
	smooth, while the 21-cm signal has some spectral structure. This property assures that the 
	foregrounds can be removed to recover the 21-cm signal. The spectral smoothness of the foregrounds 
	also suggests that they should only contaminate the low-order $k_{\parallel}$ (line-of-sight component
	of the $k$ vector) modes. However, the chromatic response of the telescope, which causes the 
	mode mixing, helps the foregrounds to contaminate the larger $k_{\parallel}$ modes. Still, the 
	foreground contamination is expected to be contained within a region (known as the ``foreground wedge")~\cite{Pober:2013jna},
	the boundary of which can be mathematically expressed as 
	\begin{equation}
		k_{\parallel}=W(z)k_{\perp}\,,
		\label{eq:wedge}
	\end{equation}
	where $W(z)$ is a $z$-dependent factor and $k_{\perp}$ is the component of the $k$ vector 
	perpendicular to the line-of-sight. 
	Eq.~\eqref{eq:wedge} also marks the ``horizon limit" when the $k_{\parallel}$ mode 
	on a given baseline corresponds to the chromatic sine wave created by a flat-spectrum 
	source of emission located at the horizon~\cite{Liu:2019awk}.
	A foreground contamination due to spatially unclustered radio sources at the horizon, 
	with a frequency-independent emission spectrum, will lie below this line and 
	we should observe the clean 21-cm signal above this line. However, due to some spectral
	features in the foregrounds, calibration error etc., foregrounds may contaminate the $k$ space
	beyond the horizon limit~\cite{Liu:2019awk}.
	Based on the above possibilities, the \texttt{21cmSense} package 
	considers three foreground contamination scenarios: ``pessimistic", ``moderate" and
	``optimistic". In the moderate scenario, the wedge is assumed to extend to 
	$\Delta k_{\parallel} = 0.1 h \mpci$ beyond the horizon wedge limit.
	In the optimistic scenario, the boundary of the foreground wedge is set by the
	FWHM of the primary beam of HERA and there is no contamination beyond this boundary. 
	Finally, in the pessimistic scenario, the foreground wedge extends $\Delta k_{\parallel} = 0.1 h \mpci$
	beyond the horizon limit, and only the instantaneously redundant baselines are combined coherently.  
	
	\subsection{Distinguishing between CDM and FDM}
	
\begin{figure*}
		\centering
		\includegraphics[width=0.9\textwidth]{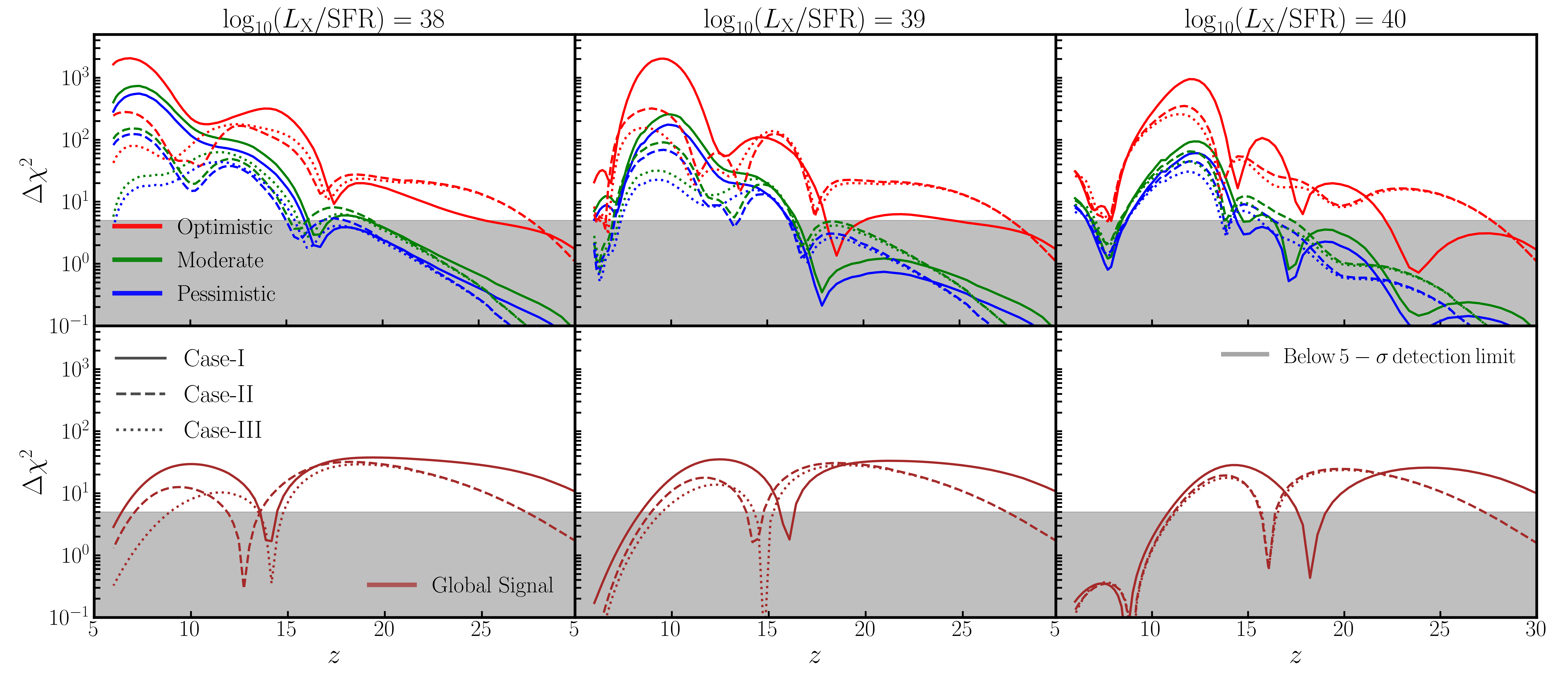}
		\caption{We show $\Delta \chi^2$, the statistical significance with which HERA can distinguish between CDM and FDM
			models, at different redshifts.  Top and bottom panels show results for power spectrum and global signal, whereas
			the different columns show results for different X-ray heating efficiencies, indicated at the top. 
			Different colors on the top panels refer to different foreground contamination scenarios.
			Solid, dashed and dotted curves 
			correspond to 	\texttt{Case-I} (no $\vcb$, no LW feedback, no additional heating),
			\texttt{Case-II} (with $\vcb$, with regular LW feedback, but no additional heating) and 
			\texttt{Case-III} (with $\vcb$, with regular LW feedback, and with additional heating), respectively. 
			The particle mass in the FDM model is $m_{\rm FDM} = 10^{-21}\,{\rm eV}$. 
			The shaded region show below $5-\sigma$ detection limit.}
		\label{fig:chi_square}
	\end{figure*}

	The possibility of discrimination between CDM and FDM models depends on the error  
	with which we shall be able to measure the 21-cm power spectrum. In order to gauge this 
	possibility, we calculate the chi-square difference $\Delta \chi^2$ which is essentially the mod of the difference
	of the power spectra between CDM and FDM divided by the expected measurement error. In 
	calculating this $\Delta \chi^2$, we assume that CDM is the correct model. We have plotted the
	$\Delta \chi^2$ calculated using the 21-cm power spectrum in the top panels of Fig.~\ref{fig:chi_square}.
	For comparison, we have also plotted the $\Delta \chi^2$ for the global signal 
	in the bottom panels of Fig.~\ref{fig:chi_square}. For global signal, we have assumed an error of 
	$5$ mK throughout the $z$ range.

	Comparing the top and bottom panels, it is evident that the power spectrum has more discriminating
	power than the global signal. The maximum $\Delta \chi^2$ is $\sim 30$ for the global signal, whereas 
	it is $\sim 2000$ for the power spectrum. For the global signal with the lowest X-ray heating, we find that
	the \texttt{Case-I} shows the maximum $\Delta \chi^2$ at most of the redshifts in comparison to other cases. 
	This is also true for other X-ray heating cases. 
	However, when we introduce relative velocity $\vcb$, LW feedback (\texttt{Case-II}) and additional
	heating (\texttt{Case-III}), we find that the overall $\Delta \chi^2$ drops, although \texttt{Case-II} and \texttt{III} show higher 
	$\Delta \chi^2$ than \texttt{Case-I} in a small $z$ range 
	around $z\sim17$ (note that this depends on X-ray heating) . Below $z\sim18$, we see some 
	difference in $\Delta \chi^2$ between \texttt{Case-II} and \texttt{III} for the lowest X-ray heating. 
	This difference fades away as the X-ray efficiency increases.

	Considering the results for the power spectrum (top panels), we see that at $z<17$, 
	$\Delta \chi^2$ values are higher for \texttt{Case-I} in all foreground contamination scenarios. 
	This changes for the optimistic foregrounds and we see that at $z>17$ \texttt{Case-II} and \texttt{Case-III}
	show higher $\Delta \chi^2$ in comparison to \texttt{Case-I}. For the moderate and pessimistic foregrounds, 
	this is true, but for a very limited redshift range. 
	From Figs.~\ref{fig:power-spectrum-k} and \ref{fig:power-spectrum-z}, 
	we see that for CDM in this range, the power spectrum amplitude is higher at small $k$
	in \texttt{Case-II} and \texttt{III} in comparison to \texttt{Case-I}. 
	The combination of this and the small error for the optimistic
	foregrounds make $\Delta \chi^2$ higher for \texttt{Case-II} and \texttt{III}. 
	Considering the optimistic foregrounds for \texttt{Case-I} with
	lowest X-ray heating, we see that the highest peak of $\Delta \chi^2$ ($\sim 2000$) 
	occurs at $z\sim7$ and the second highest peak ($\sim300$) occurs at $z\sim13$. 
	Note that we have several peaks in these $\Delta \chi^2-z$ plots, and their locations depend
	on the delay in different processes between CDM and FDM models, and also on the choice of 
	astrophysical parameters. We shall mainly focus on the first two highest peaks.
	When we introduce $\vcb$ and LW feedback effects (\texttt{Case-II}), we see that the peak values are suppressed 
	($\sim300$ for the first peak and $\sim150$ for the second). Additional heating drops the peak
	values further ($\sim70$ for the first peak and $\sim150$ for the second) and it has maximum
	effect on the first peak. The above is true for moderate and pessimistic backgrounds, only the peak values 
	change. For the moderate and high X-ray heating, we see that the peak locations and their amplitudes 
	change. Peak amplitude is lowest for the highest X-ray heating. All the discussion for the 
	lowest X-ray heating also holds for moderate and high X-ray heating. However, the effect due to the 
	additional heating decreases with increase in X-ray efficiency and is minimal for  
	$\log_{10}(L_{\rm X}/{\rm SFR})=40$. 
	
	Overall, the discussion above suggests that the presence of $\vcb$ and LW feedback (which mainly affect CDM),
	along with any heating, be it X-ray or additional (which affect both CDM and FDM), lowers our
	ability to discriminate between the CDM and FDM models.

	\begin{figure}
		\centering
		\includegraphics[width=\columnwidth]{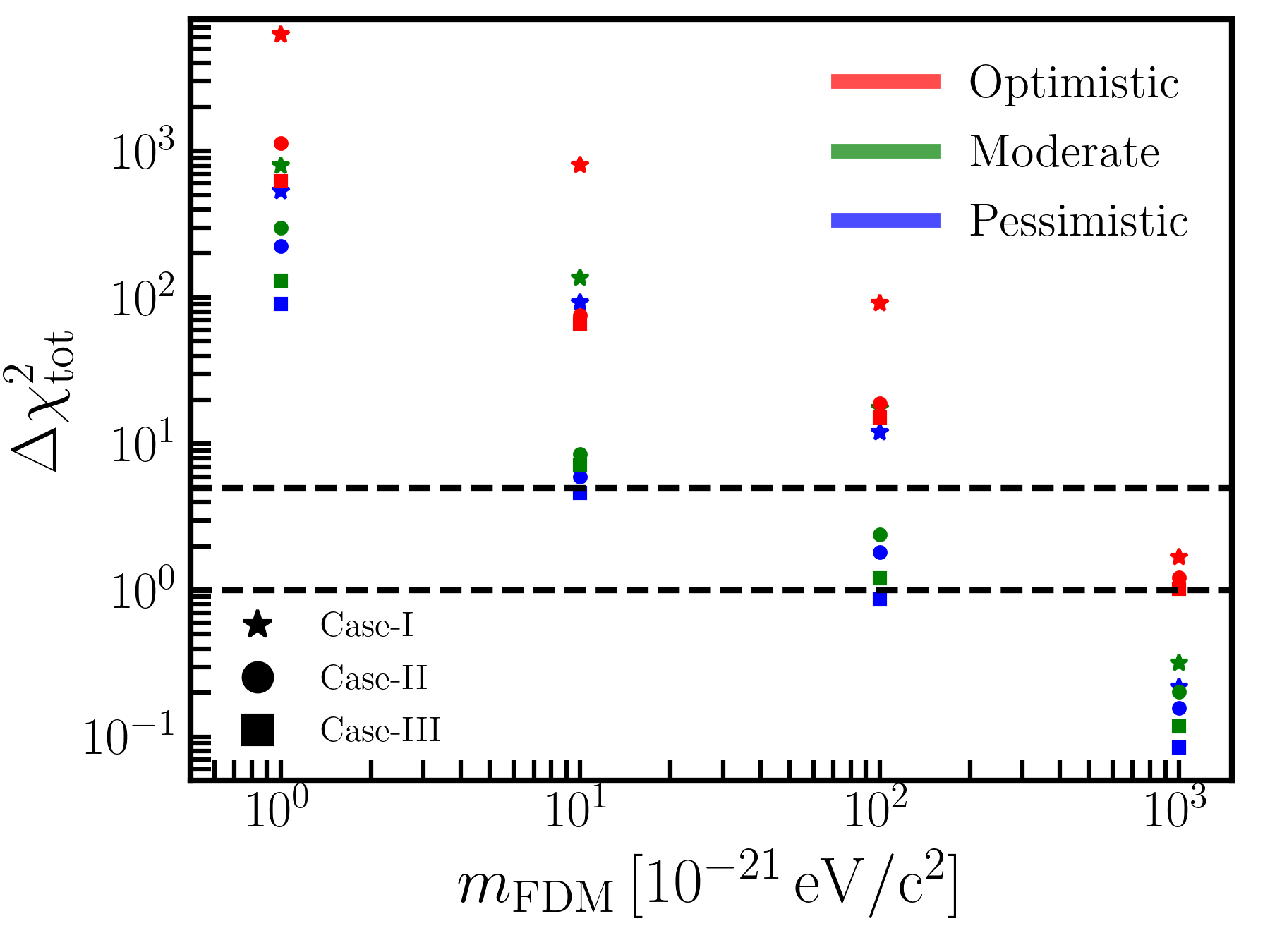}
		\caption{This shows $\Delta \chi^2_{\rm tot}$, which is the sum of $\Delta \chi^2$
		(Fig.~\ref{fig:chi_square}) over all the $z$ values, for the different values of 
		$m_{\rm FDM}$. For all the results we fix $\log_{10}(L_{\rm X}/{\rm SFR})=39$. 
		Different colours show different foreground contamination scenarios. 
		Star, circle and square markers 
		correspond to 	\texttt{Case-I} (no $\vcb$, no LW feedback, no additional heating),
		\texttt{Case-II} (with $\vcb$, with regular LW feedback, but no additional heating) and 
		\texttt{Case-III} (with $\vcb$, with regular LW feedback, and with additional heating), 
		respectively. We place two horizontal lines to show the $1-\sigma$ and $5-\sigma$ detection
		limits.}
		\label{fig:chi_square_FDM}
	\end{figure}

	Finally,  we explore the possibility of distinguishing the CDM and FDM models for different $m_{\rm FDM}$. 
	Note that, as $m_{\rm FDM}$ increases, FDM results approach CDM results. There will be a maximum 
	limit in $m_{\rm FDM}$, above which the  CDM and FDM models cannot be differentiated with HERA. 
	We investigate this limit in 
	Figure~\ref{fig:chi_square_FDM} which shows $\Delta \chi^2_{\rm tot}$, 
	the sum of $\Delta \chi^2$ (Figure~\ref{fig:chi_square})
	over all the $z$ values, for four different $m_{\rm FDM}$ values between $10^{-21}\,{\rm eV}$ and
	$10^{-18}\,{\rm eV}$. $\Delta \chi^2_{\rm tot}$ broadly determines the overall discriminating power
	of HERA. We choose two values of $\Delta \chi^2_{\rm tot}$, $1$ ($68\%$ confidence) and 
	$5$ ($99.99\%$ confidence), as metric of discriminating power. Considering the optimistic foreground
	scenario, we find that HERA is able to discriminate between CDM and FDM models (for all the three
	cases considered) up to $m_{\rm FDM} \approx 10^{-18}\,{\rm eV}$ with $68\%$ confidence and up to 
	$m_{\rm FDM} \approx 10^{-19}\,{\rm eV}$ with $99.99\%$ confidence. When we consider moderate and
	pessimistic foreground contamination, the same limits go down by an order, and HERA is able to tell apart
	the CDM and FDM models up to $m_{\rm FDM} \approx 10^{-19}\,{\rm eV}$ with $68\%$ confidence and up to 
	$m_{\rm FDM} \approx 10^{-20}\,{\rm eV}$ with $99.99\%$ confidence. Note that if we 
	consider only \texttt{Case-I}---incorporating neither $\vcb$, feedback nor additional heating---then 
	HERA is able to differentiate between CDM and FDM up to 
	$m_{\rm FDM} \approx 10^{-19}\,{\rm eV}$ with $99.99\%$ confidence even for the moderate and
	pessimistic foreground contamination scenario. This would overestimate the 21-cm $m_{\rm FDM}$ bound by an order of magnitude.

	\subsubsection{Comparison with {\tt 21cmFast} {\tt 3.1.3}}
	
	While the writing of this paper was in progress, Ref.~\cite{Munoz:2021psm} came out with the release of {\tt 21cmFast} version {\tt 3.1.3}\footnote{\href{https://github.com/21cmfast/21cmFAST}{github.com/21cmfast/21cmFAST}}. This python-based version includes the delaying effects of $v_\mathrm{cb}$ and the LW feedback, as well as the separation of Pop-II and Pop-III stars into molecular and atomic cooling halos, respectively.
	
\begin{figure}
		\centering
		\includegraphics[width=\columnwidth]{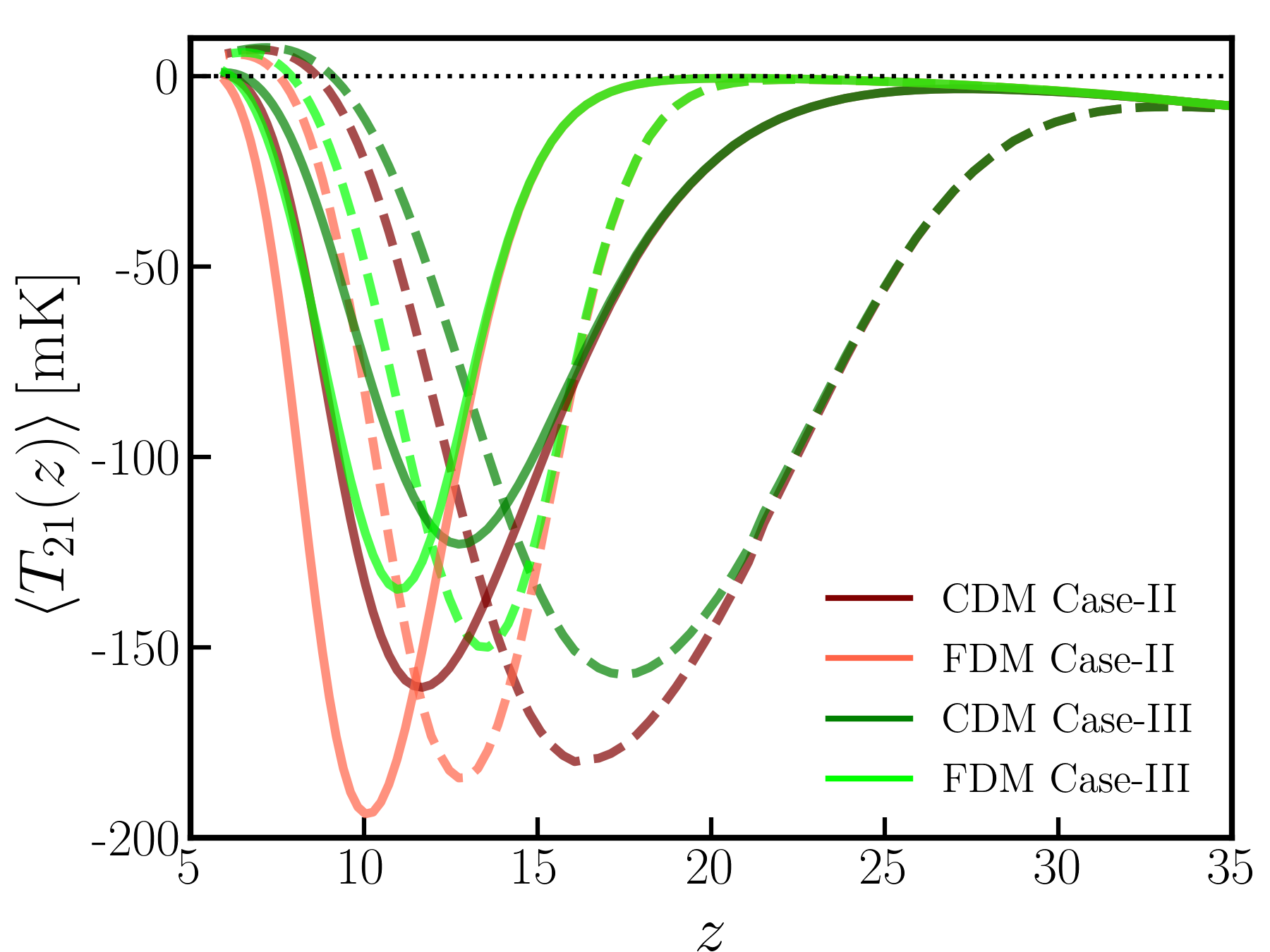}
		\caption{Comparison of the 21-cm global signals between {\tt 21cmvFast} (dashed) and {\tt 21cmFast} 3.1.3 (solid). For {\tt 21cmvFast} we use the fiducial values of Table~\ref{tab:parameters-fid}, while for {\tt 21cmFast} 3.1.3 we adopt the EOS2021 values (see Table 1 in Ref.~\cite{Munoz:2021psm}), and in both versions we set $\log_{10}(L_X/\mathrm{SFR})=39$. For the FDM cases, we set $m_\mathrm{FDM}=10^{-21}\,\mathrm{eV}$. The different cases shown here correspond to the same cases as in Fig.~\ref{fig:chi_square} and Fig.~\ref{fig:chi_square_FDM}.}
		\label{fig:versions_comparison}
	\end{figure}

	In a similar manner to our implementation in {\tt 21cmvFast}, we  have incorporated the \lya and CMB heating effects as well as the FDM transfer function in {\tt 3.1.3}. We compare the global signals of the two versions in Fig.~\ref{fig:versions_comparison}, for {\tt Case-II} and {\tt Case-III}. Most noticeably, we see that the CDM signals are significantly delayed in the new version of {\tt 3.1.3}. This is mainly due to the new modelling of the star formation rate density in {\tt 3.1.3}.

	We have also repeated the $\Delta\chi^2$ analysis in {\tt 3.1.3}. Since the CDM signal (global and power spectrum) is delayed in {\tt 3.1.3} (compared to {\tt 21cmvFast}), the difference between the CDM and FDM scenarios becomes less pronounced. However, this is compensated by the fact that the HERA sensitivity is considerably higher at lower redshifts. Overall, we find that under the modelling of {\tt 3.1.3}, the FDM scenario is more easily detectable, but none of our conclusions change by more than roughly a factor two. In particular, we find using version {\tt 3.1.3} as well that the heating effects reduce the ability of HERA to distinguish between CDM and FDM, as expected.   
	
	\subsection{Fisher Matrix Forecasts}
	
	Analyses in the previous section convince us that HERA has a high possibility to distinguish between the CDM 
	and FDM models. Now, if we consider FDM to be the true model, then it becomes important to estimate how
	well we can constrain the different parameters of the model. We use the
	Fisher matrix formalism to calculate the possible accuracy with which we can estimate 
	the FDM model parameters for the HERA observations. The Fisher matrix may be written 
	as~\cite{Jungman:1995bz, Bassett:2009tw, Jimenez:2013mga}
	\begin{equation}
		F_{\alpha,\beta} = \sum_{k,z} \frac{\partial \Delta^2_{21}(k,z)}{\partial \alpha}
		\frac{\partial \Delta^2_{21}(k,z)}{\partial \beta} \frac{1}{{\rm var} \left[ \Delta^2_{21}(k,z) \right]}\,,
	\end{equation}
	where $(\alpha,\beta)$ represent different parameters of the model and the sum runs over all the 
	$k$ modes and redshifts. Here, ${\rm var} \left[ \Delta^2_{21}(k,z) \right]$ is the expected variance for the
	observable $\Delta^2_{21}(k,z)$ which we calculate using the \texttt{21cmSense} 
	package (Section~\ref{sec:sensitivity}). Here we have assumed that the different $k$ and $z$ bins are 
	independent. 
	
	The inverse of this Fisher matrix $[F_{\alpha,\beta}]^{-1}$ gives us the covariance 
	matrix $C_{\alpha,\beta}$ for the errors in different parameters. We have considered three astrophysical
	parameters: $\log_{10}(L_{\rm X}/{\rm SFR})$ ($39$ and $40$), 
	$\zeta$ ($20$), $f^0_{\ast}$ ($0.05$), and four cosmological parameters: 
	$h$ ($0.6766$), $\Omega_{m0}$ ($0.3111$), $\Omega_{b0}$ ($0.0489$), $m_{\rm FDM}$ ($10^{-21}\,{\rm eV}$),
	with fiducial values given inside the brackets,
	for our analysis. For the FDM particle mass $m_{\rm FDM}$, we consider $\log_{10}m_{\rm FDM}$ as a 
	parameter for numerical reasons. Note that we have several other astrophysical and cosmological parameters
	(see table~\ref{tab:parameters-fid}) in our model. One should really take all the parameters into 
	consideration for any real analysis. Then the unimportant parameters can be marginalized if required. 
	However, including more parameters generally degrades the constraints. 
	Also, not all the parameters are equally important. For example, the parameters 
	$B$ (determines the LW feedback strength, and appears in eq.~(11) of ref.~\cite{Munoz:2019rhi}), 
	$V^{(0)}_{\rm cool}$ and $V^{\HI}_{\rm cool}$  are not very 
	important for the FDM analysis. FDM models suppress 
	halos much more massive than molecular cooling threshold $M_{\rm cool}$
	(eq.~(11) of ref.~\cite{Munoz:2019rhi}),
	which depends on both  $\vcb$
	and $B$. Hence, FDM models are not sensitive to both $V^{(0)}_{\rm cool}$ and $B$. 
	Similarly, $V^{\HI}_{\rm cool}$ or $M^{\HI}_{\rm cool}$ (which appear in 
	eq.~(12) of ref.~\cite{Munoz:2019rhi})
	have a weak effect on the power spectrum, as well as on the global signal. 
	Other astrophysical parameters like $\lambda_{\rm MFP}$, $\alpha_X$, $E_{\rm min}$,
	and cosmological parameters such as $\sigma_{8,0}$ or $n_s$ have some effect on the signal. However, to limit
	our discussion, we choose not to include them in our analysis.

	\begin{table*}
		\centering
		\scalebox{0.75}{
			\begin{tabular}{|c|c|c|c|c|c|c|c|c|}
				\hline
				Models & Foreground & $\log_{10}\frac{L_{\rm X}}{{\rm SFR}}$ & $\zeta$ & $f^0_{\star}$ & $h$ & $\Omega_{m0}$ & $\Omega_{b0}$ & $\log_{10} m_{\rm FDM}$ $\left( \frac{\Delta m_{\rm FDM}}{m_{\rm FDM}}\% \right)$\\ \hline
				\multirow{3}{*}{$\Delta^2_{21}$, \texttt{Case-II}, Moderate X-ray} & Optimistic & 0.0015 & 0.12 & 0.00042 & 0.0014 & 0.00088 & 0.00016 & 0.0042 ($1.0\%$)\\ \cline{2-9} 
				& Moderate & 0.0074 & 0.51 & 0.0016 & 0.0054 & 0.0011 & 0.00062 & 0.017 ($3.8\%$)\\ \cline{2-9} 
				& Pessimistic & 0.01 & 0.69 & 0.0022 & 0.0072 & 0.0017 & 0.00084 & 0.024  ($5.4\%$)\\ \hline
				
				\multirow{3}{*}{$\Delta^2_{21}$, \texttt{Case-III}, Moderate X-ray} & Optimistic & 0.0046 & 0.12 & 0.00078 & 0.002 & 0.00079 & 0.00022 & 0.006 ($1.4\%$) \\ \cline{2-9} 
				& Moderate & 0.021 & 0.52 & 0.0035 & 0.0094 & 0.0036 & 0.0011 & 0.027 ($6.03\%$) \\ \cline{2-9} 
				& Pessimistic & 0.029 & 0.73 & 0.0049 & 0.013 & 0.0051 & 0.0016 & 0.038 ($8.4\%$) \\ \hline
				
				\multirow{3}{*}{$\Delta^2_{21}$, \texttt{Case-III}, High X-ray} & Optimistic & 0.0053 & 0.077 & 0.0015 & 0.0013 & 0.00062 & 0.00019 & 0.0084 ($1.9\%$) \\ \cline{2-9} 
				& Moderate & 0.03 & 0.53 & 0.012 & 0.004 & 0.0019 & 0.0011 & 0.067 ($14.3\%$) \\ \cline{2-9} 
				& Pessimistic & 0.043 & 0.76 & 0.018 & 0.0054 & 0.0025 & 0.0015 & 0.098 ($20.2\%$) \\ \hline
				
				\multicolumn{2}{|c|}{Global, \texttt{Case-III}, Moderate X-ray}  & 0.38 & 9.15 & 0.042 & 0.062 & 0.059 & 0.008 & 0.22 ($39.7\%$) \\ \cline{1-9}
			\end{tabular}
		}
		\caption{$1-\sigma$ constraints on the parameters with HERA observations for different models.  
			The different foreground contamination scenarios are indicated in the different rows.  
			For high X-ray efficiency, there is not much difference between \texttt{Case-II} and \texttt{III}, so we
			show results only for \texttt{Case-III}. We also show results for global signal (bottom row) with an error of $5$ mK
			throughout the redshift range. 
		}
		\label{tab:constraints}
	\end{table*}

\begin{figure*}
		\centering
		\includegraphics[width=0.8\textwidth]{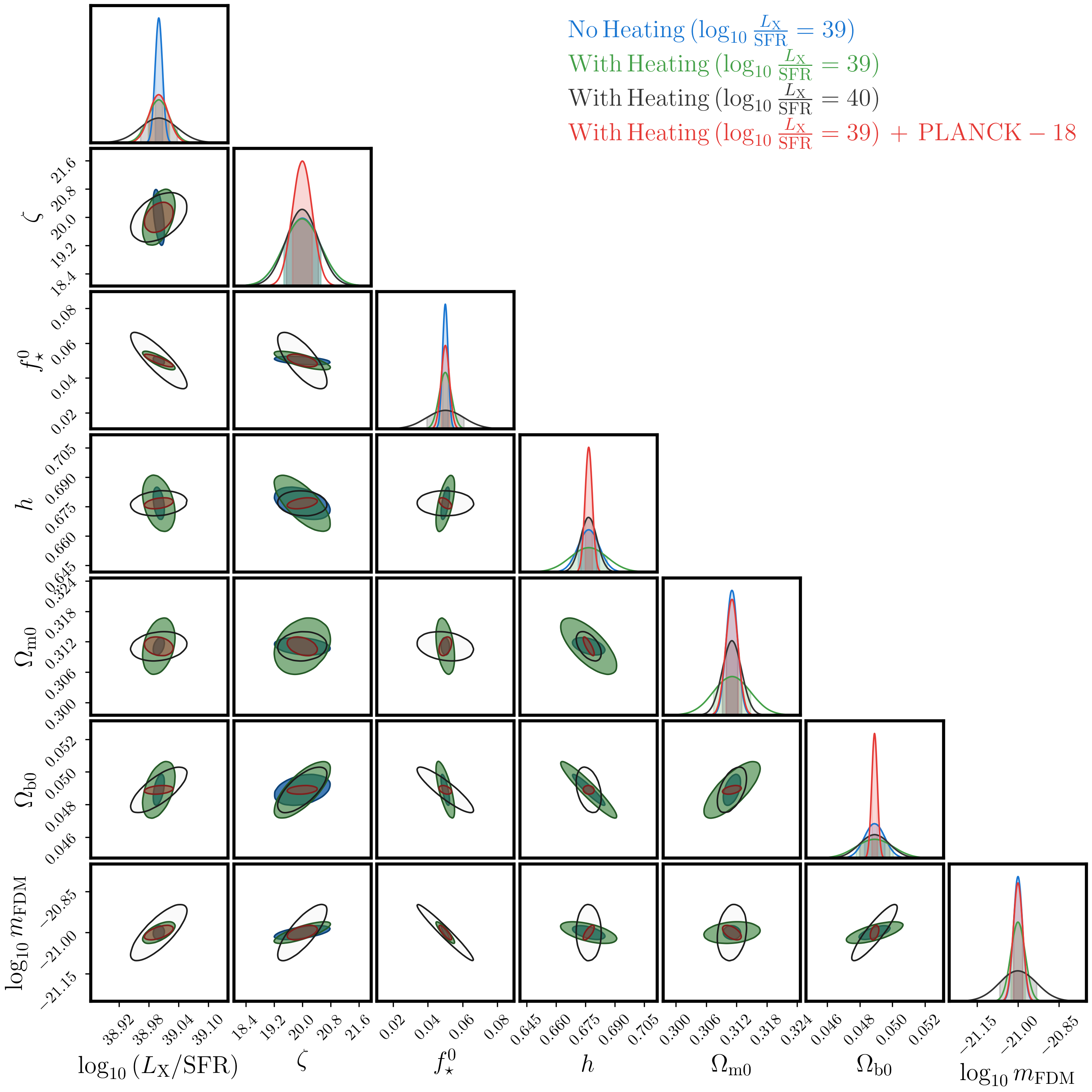}
		\caption{Effects on the parameter constraint forecasts for HERA in presence of additional
			heating. Blue and Green ellipses correspond to models without and with additional heating for a fixed
			value of X-ray efficiency $\log_{10}\frac{L_{\rm X}}{{\rm SFR}}=39$. For comparison, we have also plotted 
			results for $\log_{10}\frac{L_{\rm X}}{{\rm SFR}}=40$  with additional heating. 
			In addition, we  add to our Fisher matrix the covariance matrix calculated for \texttt{PLANCK-18}~\cite{Planck:2018vyg} CMB data using a separate MCMC analysis (see Ref.~\cite{Abadi:2020hbr} for details) and show the results in red. 
			The \texttt{PLANCK} data is added only for $\log_{10}\frac{L_{\rm X}}{{\rm SFR}}=39$  with
			additional heating. The ellipses span $1-\sigma$ confidence intervals. All the results correspond to the moderate foreground contamination scenario. Note that the black ellipse of $L_X=40$ is centered at 39, to facilitate a comparison.}
		\label{fig:fisher-heat}
	\end{figure*}

	As the FDM particle mass is an important quantity in our study, we briefly discuss the correlation of FDM particle mass with other 
	parameters. Overall, we see that it is very much degenerate with other parameters. 
	An increase in the FDM particle mass lowers the suppression effect and increases the number of
	smaller halos. This reduces the delay in structure formation, which in turn
	makes CD, the X-ray heating era and the EoR occur earlier. The same
	can be achieved by increasing $f^0_{\star}$, $\log_{10}\frac{L_{\rm X}}{{\rm SFR}}$ and $\zeta$.
	Therefore, all these parameters are expected to be negatively correlated with $\log_{10} m_{\rm FDM}$ as
	the effects of increasing $\log_{10} m_{\rm FDM}$ can be compensated by decreasing any 
	of these three parameters. However, in Fig.~\ref{fig:fisher-heat}, we find that 
	$\log_{10} m_{\rm FDM}$ is positively
	correlated with $\log_{10}\frac{L_{\rm X}}{{\rm SFR}}$ and $\zeta$. As discussed in Ref.~\cite{Jones:2021mrs},
	this is not surprising. $f^0_{\star}$, $\log_{10}\frac{L_{\rm X}}{{\rm SFR}}$ and $\zeta$ are not independent,
	but rather degenerate with each other, and also with the cosmological parameters. 
	As a result, effects due to a change in one parameter can be compensated by a combination of different
	changes in other parameters. For a particular choice of fiducial values, the
	correlation between any two parameters can be changed. However, the degeneracy between these parameters
	can be broken in the following way. $f^0_{\star}$ is important during the \lya coupling era, 
	$\log_{10}\frac{L_{\rm X}}{{\rm SFR}}$ mainly affects the X-ray heating era and $\zeta$ 
	is particularly important during Reionization. Therefore, observing these different eras separately can help 
	 break the degeneracy. Considering cosmological parameters, we see that $\log_{10}\frac{L_{\rm X}}{{\rm SFR}}$
	is positively correlated with $\Omega_{b0}$ and $\Omega_{m0}$, and negatively correlated with 
	$h$.

	Our results below 
	suggest that HERA should be able to determine the FDM particle mass to within a few percent
in the moderate foreground scenario, at $1-\sigma$ confidence. This suggests that it is possible to have a tight constraint on
	the FDM particle mass from HERA observations assuming the fiducial value of 
	$m_{\rm FDM} = 10^{-21}\,{\rm eV}$. Instead of FDM,  had we assumed CDM to be the correct model,
	 the tight constraints would indicate that the upper limit in FDM mass would be tighter than 
	$10^{-21}\,{\rm eV}$ in the CDM model. The constraints on different parameters for various
	scenarios are given in Table~\ref{tab:constraints}.

	In Fig.~\ref{fig:fisher-heat}, we show the comparison of covariance between the \texttt{Case-II},
	which has no additional heating, and \texttt{Case-III}, which includes additional heating. 
	We have shown results for moderate X-ray efficiency and with moderate foreground scenario. 
	We immediately see that the ellipses with no additional heating are smaller than the ones with 
	additional heating. This implies that the additional heating actually degrades the 
	parameter constraints. The presence of additional heating decreases the amplitude of the power
	spectrum below $z\sim17$ and this in turn reduces the SNR. However, at $z>18$, the additional heating is 
	not present and we expect to have similar results for both the cases. 
	
	A careful inspection shows that the parameter $\zeta$ is not affected much by the additional heating. 
	$\zeta$ is important during the reionization when the heating effects are generally subdominant. 
	We also see that the correlation in most of the parameters are some what different in the two cases. 
	Considering the constraints on $m_{\rm FDM}$ (Table~\ref{tab:constraints}), we see that, 
	without additional heating (\texttt{Case-II})
	HERA should be able to determine the FDM particle mass to within
	$1\%$, $3.8\%$ and $5.4\%$ in the optimistic, moderate, and pessimistic foreground scenarios, 
	respectively, at $1-\sigma$ confidence. This is  a factor of $\sim1.5$ better than 
	in \texttt{Case-III}. 
	
	For comparison, we have also shown results for \texttt{Case-III} with high X-ray efficiency.
	We see that the constraints on the most of the parameters degrade for high X-ray efficiency, 
	except for $h$ and $\Omega_{m0}$. Constraints on $h$ and $\Omega_{m0}$ are better for 
	high X-ray efficiency. Also, $h$ and $\Omega_{m0}$ show mild correlation with $\log_{10} m_{\rm FDM}$.
	In the case of high X-ray, HERA should be able to determine the FDM particle mass to within
	$14.3\%$ in the moderate foreground scenario at $1-\sigma$ confidence. The same is $1.9\%$ and $20.2\%$
	in optimistic and pessimistic foreground scenarios, which we do not show here 
	(see Table~\ref{tab:constraints}). Note that, for high X-ray, the additional heating effect is almost
	negligible and there is not much difference in results between \texttt{Case-II} and \texttt{Case-III}.  
	Hence we do not show the results of \texttt{Case-II} for high X-ray efficiency.

	We have also studied the implications of adding the \texttt{PLANCK-18}~\cite{Planck:2018vyg} covariance matrix on the astrophysical parameters. 
	We use a dedicated MCMC analysis to derive the covariance matrix from \texttt{PLANCK-18} CMB data\footnote{We thank Tal Abadi for performing this analysis (as in Ref.~\cite{Abadi:2020hbr}).}, and add it to our  21-cm Fisher matrix. In the future, thanks to the direct interface to {\tt CLASS}, our code will enable joint MCMC 21-cm-CMB real-data analyses to test the standard cosmological model or any extensions to it.  
	
	As expected, \texttt{PLANCK} measures the 
	cosmological parameters with great precision. Here, the idea is to check the improvement in the constrains
	on the astrophysical parameters when the uncertainty in the cosmological parameters is 
	minimized. Generally tight priors on a set of parameters help in reducing the error in other 
	parameters. Indeed we see exactly this. We add the \texttt{PLANCK} information only to the \texttt{Case-III} with moderate X-ray.
	We see that the priors result in a significant improvement and the results are comparable to that
	with \texttt{Case-II}. The constraint on  $\log_{10} m_{\rm FDM}$ is very close to that obtained for \texttt{Case-II}.

	Lastly, mostly as a fact-check, we have  plotted the covariance ellipses for the global signal at moderate X-ray, in
	comparison to the same ellipses for the power spectrum (not shown in this paper). 
	We find that the ellipses for the power spectrum look
	minuscule in comparison to the global signal ellipses ---  we do not have very good constraint 
	on any variable for the global signal, which is expected. The global signal lacks the information
	on the different length scales which the power spectrum possesses. This gives the power spectrum 
	more constraining power. Nevertheless, a global signal experiment with an overall redshift-independent
	sensitivity of $5$ mK should be able to determine the FDM particle mass to within
	$39.7\%$ at $1-\sigma$ confidence. This error, of course, can be minimized by applying 
	priors on the different parameters.

	\section{Summary and Conclusions}
	\label{sec:conclusions}
	
	We have simulated the effects of FDM on the 21-cm global signal and power spectrum. Extending previous works, we  
	incorporated  in our simulations several important effects on the 21-cm signal, such as the DM-baryon relative velocity $\vcb$, LW radiative feedback,
	and CMB and \lya heating, and studied their impact on the 21-cm signal. For completeness, 
	we also included the results for CDM with all the effects and compared those with the FDM results. 
	
	The suppression in the number of small halos in the FDM model delays the onset of CD, 
	the ensuing epoch of heating and also the EoR, in comparison to CDM. The signature of this delay
	can be seen clearly in the global 21-cm signal, as well as in the 21-cm power spectrum. 
	FDM also influences the spatial structure of the 21-cm fluctuations. The absence of the small halos 
	in the FDM model makes the ionizing sources more biased in comparison to CDM model. 
	The DM-baryon relative velocity $\vcb$ and LW radiative feedback also delay the different
	epochs. However, these only affect CDM models, as the length scale below which the 
	halos are suppressed in FDM is well above the effective Jeans scales of both 
	$\vcb$ and LW feedback. 
	
	The additional heating, which is a combination of 
	CMB and \lya heating in our analysis, affects both FDM and CDM models. Additional heating
	increases the minimum value of the absorption signal and shifts the redshift of the minimum. 
	Additional heating also alters the amplitude of the power spectrum. 
	Also, the effect of additional heating is maximal for the FDM models, compared to CDM.
	This effect, however, is important only when the X-ray heating is not very efficient.

	We have investigated the prospects to distinguish between the CDM and FDM models
	(for fixed FDM particle mass $m_{\rm FDM} = 10^{-21}\,{\rm eV}$)
	by means of 
	$\Delta \chi^2$ values, considering both the global signal and fluctuations. 
	For global signal experiments, we have considered a $5$ mK uncertainty throughout the $z$ range.
	For the fluctuations, we have considered the future HERA observations with three foreground contamination
	scenarios: optimistic, moderate and pessimistic. We find that the power spectrum is far  superior
	(higher $\Delta \chi^2$ values) to the global signal in differentiating the CDM and FDM models. 
	However, $\Delta \chi^2$ values drop as we introduce $\vcb$, LW feedback or additional heating.
	The $\Delta \chi^2$ values also vary with X-ray heating efficiency, and we see the lowest $\Delta \chi^2$
	for the highest X-ray efficiency. Therefore, all the additional effects, like $\vcb$, LW feedback,
	heating, lower our ability to discriminate between CDM and FDM models.

	Considering different $m_{\rm FDM}$ values in the range $10^{-21}\,{\rm eV}$ to $10^{-18}\,{\rm eV}$,
	we find that HERA is able to distinguish between the CDM and FDM models up to 
	$m_{\rm FDM} \approx 10^{-18}\,{\rm eV}$ ($10^{-19}\,{\rm eV}$) with $1-\sigma$ ($5-\sigma$)
	confidence in optimistic foreground scenario, and 
	$m_{\rm FDM} \approx 10^{-19}\,{\rm eV}$ ($10^{-20}\,{\rm eV}$) with $1-\sigma$ ($5-\sigma$)
	in both moderate and pessimistic foreground scenario.

	We have also shown Fisher matrix forecasts for the 21-cm power spectrum. We have considered three astrophysical
	parameters and four cosmological parameters, including the FDM particle mass $m_{\rm FDM} = 10^{-21}\,{\rm eV}$. 
	We find that $m_{\rm FDM}$ is correlated with astrophysical parameters, as well as cosmological parameters. 
	In addition, the astrophysical parameters are themselves correlated with each other. This correlation, however, can be broken
	by observing the signal form different epochs. For moderate X-ray heating efficiency 
	($\log_{10}\frac{L_{\rm X}}{{\rm SFR}}=39$) and in the presence of additional heating, 
	we find that HERA should be able to determine the FDM particle mass to within 
	$1.4\%$, $6.03\%$ and $8.4\%$ in the optimistic, moderate, and pessimistic foreground scenarios, 
	respectively, at $1-\sigma$ confidence. These constraints improve by a factor of $\sim 1.5$, if
	we do not consider the additional heating. The constraints degrade when we consider high
	X-ray heating efficiency. 
	In addition, we find that the global signal provides a much poorer, $\mathcal{O}(1)$
	constraint on FDM particle mass. We also find that the uncertainty in the
	astrophysical parameters, as well as in $m_{\rm FDM}$, can be lowered by incorporating  \texttt{PLANCK-18}~\cite{Planck:2018vyg}
	CMB constraints on the cosmological parameters.

	A number of independent cosmological probes can be utilized to distinguish between CDM and FDM models and,
	perhaps, measure the FDM particle mass. These include CMB multipoles \cite{Planck:2019nip, Planck:2018vyg}, 
	X-ray observations \cite{Maleki:2019xya}, 
	\lya effective opacity \cite{Sarkar:2021pqh}
	and galaxy power spectrum \cite{DES:2017tss}. 
	Future low-$z$ 21-cm line-intensity mapping surveys are sensitive to FDM masses up to 
	$10^{-22}\,{\rm eV}$ or below \cite{Bauer:2020zsj}. Upcoming ``high definition CMB" experiments show a similar level of
	sensitivity \cite{Hlozek:2016lzm}. 
	Measurements of the cluster pairwise velocity dispersion using the kinetic Sunyaev-Zel’dovich (kSZ) 
	effect could reach the FDM mass limit as low as $\sim 10^{-27}\,{\rm eV}$ \cite{Farren:2021jcd}. 
	Future pulsar timing array measurements 
	could probe FDM masses around $\sim 10^{-22}\,{\rm eV}$ \cite{Porayko:2018sfa}. 
	Ref~\cite{Dentler:2021zij} found a lower limit $m_{\rm FDM}>10^{-23}\,{\rm eV}$ using a combination of
	Dark Energy Survey Year 1 Data and CMB measurements.
	Heating of the Milky Way disk leads to the limit $m_{\rm FDM} \gtrsim 10^{-22}\,{\rm eV}$ \cite{Church:2018sro}.
	Recently, Ref.~\cite{Unal:2020jiy} obtained the limits  
	$10^{-21}\,{\rm eV} <  m_{\rm FDM}  <  10^{-17}\,{\rm eV}$ 
	from measuring the mass and spin of accreting and jetted black holes 
	by analyzing their electromagnetic spectra.
	Currently, the best 
	conservative ($2-\sigma$) lower limit $m_{\rm FDM}>2\times10^{-21}\,{\rm eV}$ comes from the 
	\lya forest observations. Observations of Eridanus-II star cluster rule out the range 
	$m_{\rm FDM}=10^{-20}-10^{-19}\,{\rm eV}$~\cite{Marsh:2018zyw}. Interestingly, the combination of these bounds  leaves a small window between 
	$m_{\rm FDM}=10^{-21}-10^{-20}\,{\rm eV}$ which can potentially be probed by  future 21-cm
	intensity mapping experiments.

	There are some physical effects we chose to neglect but could be important for the 21-cm signal.
	Recently, 
	Ref.~\cite{Reis:2021nqf} have introduced a new effect in the 21-cm calculations which is 
	the multiple scattering of \lya photons in the IGM. The multiple scattering actually 
	reduces the effective distance which \lya photons can travel. 
	This has some important 
	consequences and can be important in scenarios with low X-ray efficiency. 
	In addition, due to the limitations of {\tt 21cmvFast}, we ignored the first population (Pop III) of stars, and assumed that all the stellar radiation in our simulation originated in the metal enriched second generation (Pop II), an assumption that could be relaxed with the new python-based version of {\tt 21cmFast}~\cite{Munoz:2021psm}. 
	Finally, a full treatment for both star populations would feature the delaying due to the transition time 
	from pop III to pop II stars, which depends on the properties of Pop III stars, such as their initial mass 
	function and efficiency of star formation~\cite{Cohen:2015qta, Mirocha:2017xxz, Tanaka:2018ozk,Tanaka:2020ecy, Schauer:2019ihk, Magg:2021jyc}. We leave such corrections to future work.
	
	It is important to bear in mind that all parameter constraints depend on our ability to mitigate the
	foregrounds and systematics in  future 21-cm experiments~\cite{Zhang:2015mga, Makinen:2020gvh, Cunnington:2020njn}. 
	However, looking at the 
	reasonably good constraints even in the pessimistic foreground contamination case, 
	we are hopeful that  future 21-cm experiments will be able to determine the FDM 
	particle mass with good accuracy. On the other hand, if CDM is the true model,
	we expect to rule out FDM with great confidence. Finally, we emphasize that our analysis is limited only to 
	the power spectrum, which contains only a small amount of the total information embedded in the
	21-cm field. Focusing on the velocity acoustic oscillations~\cite{Munoz:2019rhi} in the 21-cm signal, Ref.~\cite{Hotinli:2021vxg} found that upcoming 21-cm surveys could be able 
	to measure the FDM particle mass as high as $10^{-18}\,{\rm eV}$, which would be interesting to revisit under the inclusion of additional heating effects. 
    The higher order statistics, like the bispectrum~\cite{Shimabukuro:2015iqa, Watkinson:2018efd, Sarkar:2019ojl, Hutter:2019yta}, 
    trispectrum \cite{Shaw:2019qin} {\it etc.}, if measured
	with high accuracy, will provide us with more constraining power.  We leave these for  future
	study.

	\begin{acknowledgements}
We thank Julian Mu\~noz for useful discussions and comments on the manuscript.  EDK acknowledges support from an Azrieli faculty fellowship.  JF is supported by a High-Tech fellowship awarded by the BGU Kreitmann School.
\end{acknowledgements}

\end{document}